\documentclass[twocolumn,traditabstract]{aa} 
\usepackage{graphicx}
\usepackage{txfonts}
\usepackage{natbib}

\newcommand{\kms}{\hbox{km s$^{-1}$}}

\begin{document}

\title{Water in star-forming regions with Herschel: highly excited molecular emission from the NGC 1333 IRAS 4B outflow
\thanks{Herschel is an ESA space observatory with science instruments provided
by European-led Principal Investigator consortia and with important
participation from NASA.}}

\author{Gregory J. Herczeg\inst{1,2}, Agata Karska\inst{1}, Simon Bruderer\inst{1},  Lars E. Kristensen\inst{3},\\
Ewine F. van Dishoeck\inst{1,3},
  Jes K. J{\o}rgensen\inst{4}, Ruud Visser\inst{5}, Susanne F. Wampfler\inst{6}, Edwin A. Bergin\inst{5},\\ Umut A. Y{\i}ld{\i}z\inst{3},
 Klaus M. Pontoppidan\inst{7}, Javier Gracia-Carpio\inst{1}}

\institute{$^1$Max-Planck-Institut f\"ur extraterrestriche Physik, Postfach 1312, 85741 Garching, Germany; gherczeg1@gmail.com;\\
$^2$Kavli Institute for Astronomy and Astrophysics, Ye He Yuan Lu 5, Beijing, 100871, P.R. China;\\
$^3$Sterrewacht Leiden, Leiden University, P.O. Box 9513, 2300 RA Leiden, The Netherlands;\\
$^4$Niels Bohr Institute and Centre for Star and Planet Formation, University of Copenhagen, Juliane Maries Vej 30, DK-2100 Copenhagen {\O}., Denmark.\\
 $^5$Department of Astronomy, The University of Michigan, 500 Church Street, Ann Arbor, MI 48109-1042, USA\\
$^6$Institute for Astronomy, ETH Zurich, 8093 Zurich, Switzerland\\
$^7$Space Telescope Science Institute, 3700 San Martin Drive, Baltimore, MD 21218}

 \date{Received 18 August 2011; }
 \authorrunning{Herczeg et al.}
  \titlerunning{Warm water in a Class 0 outflow}

\abstract{
During the embedded phase of pre-main sequence stellar evolution, a disk forms from the dense envelope while an accretion-driven outflow carves out a cavity within the envelope.  Highly excited ($E^{\prime}=1000-3000$~K) H$_2$O emission in spatially unresolved {\it Spitzer}/IRS spectra of a low-mass Class 0 object, NGC 1333 IRAS 4B, has previously been attributed to the envelope-disk accretion shock.  However, the highly excited H$_2$O emission could instead be produced in an outflow.  As part of the survey of low-mass sources in the {\it Water in Star Forming Regions with Herschel} (WISH-LM) program, we used {\it Herschel}/PACS to obtain a far-IR spectrum and several Nyquist-sampled spectral images to determine the origin of excited H$_2$O emission from NGC 1333 IRAS 4B.  The spectrum has high signal-to-noise in a rich forest of H$_2$O, CO, and OH lines, providing a near-complete census of far-IR molecular emission 
from a Class 0 protostar.  The excitation diagrams for the three molecules all require fits with two excitation temperatures, indicating the presence of two physical components.
The highly excited component of H$_2$O emission is characterized by subthermal excitation of $\sim 1500$ K gas with a density of $\sim3\times10^{6}$ cm$^{-3}$, conditions that also reproduce the mid-IR H$_2$O emission detected by {\it Spitzer}.  On the other hand, a high density, low temperature gas can reproduce the H$_2$O spectrum observed by {\it Spitzer} but underpredicts the H$_2$O lines seen by {\it Herschel}.  Nyquist-sampled spectral maps of several lines show two spatial components of H$_2$O emission, one centered at $\sim 5^{\prime\prime}$ (1200 AU) south of the central source at the position of the blueshifted outflow lobe and a second centered on-source.  The redshifted outflow lobe is likely completely obscured, even in the far-IR, by the optically thick envelope.  Both spatial components of the far-IR H$_2$O emission are consistent with emission from the outflow.   In the blueshifted outflow lobe over 90\% of the gas-phase O  is molecular, with H$_2$O twice as abundant than CO and 10 times more abundant than OH.  The gas cooling from the IRAS 4B envelope cavity walls is dominated by far-IR H$_2$O emission, in contrast to stronger [O I] and CO cooling from more evolved protostars.  The high H$_2$O luminosity may indicate that the shock-heated outflow is shielded from UV radiation produced by the star and at the bow shock.}



\maketitle

%
%

\section{INTRODUCTION}

During the embedded phase of pre-main sequence stellar evolution, the protoplanetary disk forms out of a dense molecular envelope \citep[e.g.][]{Terebey1984,Adams1987}.  
Meanwhile, as the protostar builds up most of its mass, it drives powerful, collimated outflows into the dense envelope \citep[e.g.][]{Bontemps1996}.  These processes together eventually cause the envelope to dissipate and set the initial conditions for disk evolution and planet formation.

At the interfaces between the outflow and envelope and between envelope and disk, shocks can heat the gas and potentially produce detectable emission.  The well-studied outflow-envelope interactions produce an outflow cavity with walls heated by shocks and energetic radiation from the central star  \citep[e.g.][]{Snell1980,Spaans1995,Arce2006,vanKempen2009,Tobin2010}. 
On the other hand, observational evidence for the disk-envelope interactions has been sparse.  \citet{Velusamy2002} detected methanol emission from L1157 on scales of $\sim 1000$ AU, spatially-extended beyond the point-like continuum emission, and argued that the kinematics suggest that the emission is produced at the disk/envelope interface.  
 \citet{Watson2007} detected emission in highly excited ($E^\prime=1000-3000$~K) H$_2$O lines from the NGC 1333 IRAS 4B system and attributed the heating to an envelope-disk accretion shock within 100 AU of the star.  These observations offer two different interpretations for disk-envelope interactions, with material either entering the disk on large scales \citep{Visser2009,Vorobyov2010} or raining onto the disk at small radii \citep{Whitney1993}.
However, a persistent complication in interpreting spatially and spectrally unresolved emission as coming from a compact disk-like structure is that outflows can also produce bright emission in highly excited lines.  Molecular emission is a dominant coolant of outflow-envelope interactions, and H$_2$O emission is particularly sensitive to shocks in outflows \citep[e.g.][]{Nisini2002,vanKempen2010,Nisini2010}.

The NGC 1333 IRAS 4B system (hereafter IRAS 4B) is a Class 0 YSO \citep[$d=235$~pc,][]{Hirota2008} 
with a 0.24 M$_\odot$ disk that is deeply embedded ($A_V\sim1000$ mag) within a $2.9$ M$_\odot$ envelope (J\o rgensen et al.~2002, 2009).  Compact outflow emission is detected in many sub-millimeter (sub-mm) molecular lines \citep{difrancesco2001,Jorgensen2007}.   
Near-IR emission in all four {\it Spitzer}/IRAC bands (3.5, 4.5, 5.8, and 8.0 $\mu$m) is located in the blueshifted outflow 
lobe (J\o rgensen et al.~2006, see also Choi et al.~2011), offset by $\sim 6^{\prime\prime}$ south of the peak of interferometric sub-mm continuum 
emission.  The redshifted 
outflow is seen in sub-mm line emission \citep{Jorgensen2007} but is not detected in the near- or mid-IR because 
the outflow is located behind IRAS 4B and hidden by the high extinction of the envelope.  { In the near-IR, the {\it Spitzer}/IRAC photometry is dominated by H$_2$ emission (Arnold et al.~2011; see also Neufeld et al.~2008), with some contribution of CO fundamental emission to the 4.5 $\mu$m bandpass (Tappe et al.~2011; see also Herczeg et al.~2011).}
Excited water emission 
from the NGC 1333 IRAS 4 system, including both IRAS 4A and 4B, was detected with {\it ISO}/LWS, but with too low spatial resolution to attribute 
the emission to any component in the system \citep{Giannini2001}.  { H$_2$O maser emission has also been seen from dense gas associated with the IRAS 4B outflow, although with a different position angle than the molecular outflow \citep[]{Rodriguez2002,Furuya2003,Marvel2008,Desmurs2009}.}

\begin{figure}[!t]
\includegraphics[width=90mm]{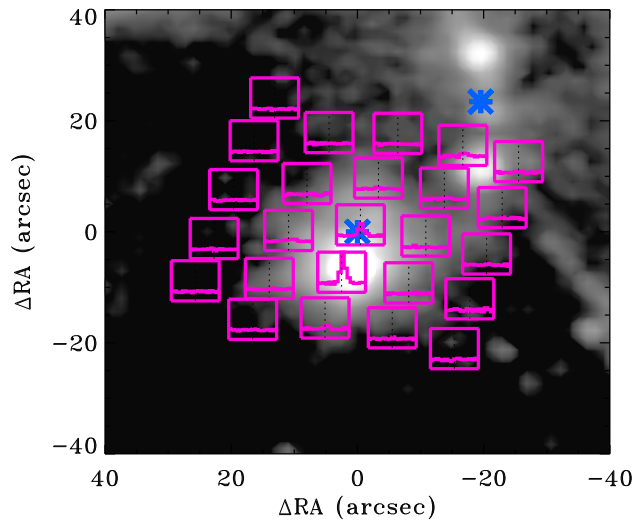}
\caption{The location of the $5\times5$ spaxel array (purple spectral map) against an 4.5 $\mu$m image obtained with {\it Spitzer}/IRAC (grayscale) from \citet{Jorgensen2006}.  The spectral map shows that the continuum-subtracted emission in the o-H$_2$O $6_{16}-5_{05}$ 82.03 $\mu$m ($E^\prime=643$ K) line is produced mostly in the blueshifted outflow lobe.  The sub-mm positions of IRAS 4B and IRAS 4A are marked with blue asterisks.}
\label{fig:spaxmap.ps}
\end{figure}

In a sample of low-resolution {\it Spitzer}-/IRS spectra of 30 Class 0 objects, \citet{Watson2007} found water emission from highly-excited levels in only IRAS 4B.  The lines have upper levels with high excitations ($1000-3000$ K) and high critical densities ($\sim 10^{11}$ cm$^{-3}$).  Watson et al.~inferred that the emitting gas has a high density and argued that the high density indicates that the emission is produced in a $\sim 2$~\kms\ accretion shock at the envelope-disk interface.  They also argued that the emission was detected only from IRAS 4B because the viewing angle may be well-aligned with the outflow, allowing a clear view of the embedded disk, and because the timescale for such high envelope-disk accretion rates may be short.  
H$_2$O emission has since been detected in at least two other components of the IRAS 4B system:  (1) narrow ($\sim 1$~\kms, with rotation signatures) p-H$_2^{18}$O $3_{13}-2_{20}$ ($E^\prime=204$ K) emission in a spatially compact region, likely a (pseudo)-disk of $\sim 25$ AU in radius \citep{Jorgensen2010}; and (2)  broad (FWHM$\sim24$ \kms), spatially unresolved emission from low-excitation H$_2$O ($E^\prime=50-250$ K) lines, which are consistent with an outflow origin 
 \citep{Kristensen2010} but too broad for the $\sim 2$~\kms\ velocity expected of envelope gas in free-fall striking the disk at 25--100 AU \citep[e.g.][]{Shu1977}.

The highly-excited mid-IR emission and the broad line profiles of lower-excitation lines could be reconciled if either (a) the high- and low-excitation H$_2$O line emission originates in different locations, if (b) models for the envelope-disk accretion shock underpredict line widths, or if (c) both the high- and low-excitation H$_2$O lines are produced in the outflow.  In this paper, we analyze a {\it Herschel}/PACS far-IR spectral survey and spectral imaging of IRAS 4B to resolve the discrepancy in the different possible origins of H$_2$O emission from IRAS 4B.  The far-IR spectrum of IRAS 4B is as rich in lines as its mid-IR {\it Spitzer}/IRS spectrum.  In Nyquist-sampled maps, the H$_2$O emission is spatially offset from the continuum emission in the direction of the blueshifted outflow.
The excitation of warm H$_2$O lines detected with {\it Herschel}/PACS and with {\it Spitzer}/IRS together can be explained by emission from an isothermal slab.  An additional physical component(s) is located on-source and likely traces material closer to the base of the outflow.  { No evidence is seen for an envelope-disk accretion shock.  This result is consistent with a conclusion by \citet{Tappe2011}, obtained contemporaneous to the results in this paper, that the H$_2$O emission seen in the {\it Spitzer}/IRS spectra also coincide with the southern outflow position.}
We discuss the implications of these results for outflows and for the prospects of observationally studying disk formation with H$_2$O lines.

\section{OBSERVATIONS AND DATA REDUCTION}

We obtained far-IR spectra of  NGC 1333 IRAS 4B \citep[03$^h$29$^m$12$^s$.0 +31$^\circ$13$^\prime$08$\farcs$1; ][]{Jorgensen2007} on 15--16 March 2011 with the PACS instrument on {\it Herschel} \citep{Pilbratt2010,Poglitsch2010} as part of the WISH Key Program \citep{vanDishoeck2011}.  The observations presented here consist of a complete scan of the 52--208 $\mu$m spectral range and deep,  Nyquist-sampled spectral maps in four narrow ($\sim 0.5-1$ $\mu$m) wavelength regions.  Each PACS spectrum includes observations of two different nod positions 
located at $\pm3^{\prime}$ from the science observation to subtract the instrumental background.   All PACS data were reduced with HIPEv6.1 \citep{Ott2010}. 
We supplemented the PACS observations with re-reduced archival {\it Spitzer}/IRS spectra that were previously analyzed by \citet{Watson2007}.  Details of the observations and reduction are described in the following subsections.

\begin{figure}[!t]
\includegraphics[width=90mm]{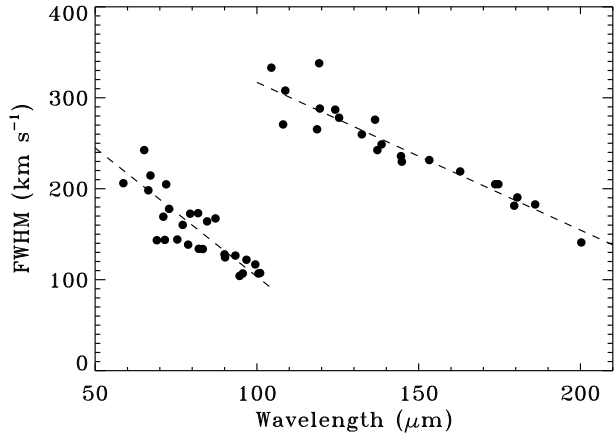}
\vspace{5mm}
\caption{The FWHM from initial fits to strong lines in our full scan of the 52--208 $\mu$m wavelength range.  The lines are spectrally unresolved and provide a measure of the spectral resolution.  The line widths for our final fits were set by linear fits to the FWHM at $>$100 and $<$100 $\mu$m.}
\label{fig:specres.ps}
\end{figure}

\subsection{Complete {\it Herschel}/PACS far-IR spectrum}

{ The 52--208 $\mu$m spectrum of IRAS 4B was obtained in 2.7 hrs of integration with PACS.}  PACS observed IRAS 4B simultaneously in the first order $>100$ $\mu$m and in the second order at $<100$ $\mu$m.  The grating resolution varies between $R=1000-2000$ at $>100$ $\mu$m and $R=3000-4000$ at $<100$ $\mu$m.  Large grating steps are used, so the spectrum is binned to $\sim 2$ pixels per resolution element.   A spectral flatfielding was applied to the data to improve the S/N.   The observed flux was normalized to the telescopic background and subsequently calibrated from observations of Neptune, which is used as a spectral standard.   { Appendix A describes a new method to calibrate the flux in PACS spectra between 97--103 $\mu$m and above $>190$ $\mu$m.}  The relative flux calibration is accurate to $\sim 20\%$ across { most of} the spectrum.

The spectral scan produced a single $5\times5$ spectral map over a $47^{\prime\prime}\times47^{\prime\prime}$ field-of-view.  
The central spaxel is centered at the location of the sub-mm continuum peak.  An adjacent spaxel located $9\farcs4$ to the SE (PA=249$^\circ$) is centered on the blueshifted outflow position (Fig.~\ref{fig:spaxmap.ps}).   Most of the line and continuum flux is located in these two spaxels.   The southern outflow of NGC 1333 IRAS 4A is located in the NW corner of the array.

Extracting line fluxes requires an assessment of the distribution of flux on the detector caused by both the point-spread function of Herschel and the spatial extent of the emission.
For a well-centered point source, the encircled energy of in a single spaxel is $\sim$70\% at $\leq 100$ $\mu$m wavelengths and declines to $\sim 40$\% near 200 $\mu$m.  However, the signal-to-noise decreases if the spectrum is extracted from many spaxels.
 The emission line spectrum is obtained by adding the flux from the central spaxel and the outflow spaxel.  The line fluxes are subequently corrected for light leakage and the spatial extent in the emission by comparing line fluxes from this 2-spaxel extraction with the line fluxes extracted from a $3\times3$ spaxel area centered on the central spaxel.  The line flux ratio between the 2-spaxel and $3\times3$ spaxel extraction was calculated for strong lines of all detected molecules.  The fluxes from the 2-spaxel extraction are then divided by a wavelength-dependent correction that is $0.76$ at $<100$ $\mu$m and then decreases linearly to 0.6 at 180 $\mu$m.  
The wavelength dependence of this correction includes both the point-spread function of Herschel and the wavelength dependence in the spatial distribution of the detected emission.  Finally, high signal-to-noise PACS spectra of the point source HD 100546 \citep{Sturm2010} were then used to correct for emission leaked beyond the $3\times3$ spaxel area.  This approach assumes a similar spatial distribution for all molecules and for lines at nearby wavelengths, and introduces a $\sim 10\%$ uncertainty in relative fluxes.  The overall uncertainty in flux calibration is $\sim 30\%$.  The typical noise level in the two-spaxel extraction ($\sim 9.4\times18.8^{\prime\prime}$) is $\sim 0.4$ Jy per resolution element in the extracted continuum spectrum.

\begin{table*}[!b]
\caption{Location of PACS line and continuum emission from the IRAS 4B system}
\begin{tabular}{|lcccccccc|}
\hline
& Species & Line  & $E_{up}$ (K) & $\lambda$ ($\mu$m) & $\Delta$RA ($^{\prime\prime}$) & $\Delta$Dec  ($^{\prime\prime}$) & $\sigma$(RA)  ($^{\prime\prime}$) & $\sigma$(Dec)  ($^{\prime\prime}$) \\
\hline
& CO  &49--48                    &  6457& 53.9 &    $3.4\pm1.5$    & $-2.4\pm0.5$    & $7\pm2$ & $9.9\pm1.0$ \\
& o-H$_2$O &$5_{32}-5_{05}$ &732& 54.5 &   $1.0\pm1.0$     & $-5.4\pm0.5$   & $9.9\pm1.0$  & $10.4\pm1.0$\\
&[O I] &$^3P_1-^3P_2$         &228 & 63.2  &  $-1.3\pm0.3$   & $-3.9\pm0.2$   & $11.8\pm0.7$ & $12.2\pm0.8$\\
& o-H$_2$O &$8_{18}-7_{07}$&1070 &63.3  &  $0.2\pm0.2$  &  $-5.1\pm0.3$      & $11.2\pm0.3$ & $12.2\pm0.4$\\
& o-H$_2$O &$8_{08}-7_{17}$ &1070&63.5  &   $0.2\pm0.3$ &  $-5.2\pm0.4$      & $11.3\pm0.6$ & $14.4\pm1.0$\\
& \multicolumn{2}{c}{Continuum}                    &--   & 63.3  & $0.4\pm0.4$ &  $-0.7\pm0.5$      & $11.5\pm0.5$ & $10.1\pm0.8$\\
& o-H$_2$O &$2_{21}-1_{10}$&194 &108.1  &  $0.3\pm0.3$  &  $-3.1\pm0.2$    & $12.5\pm0.4$   &    $18.1\pm0.4$ \\
& CO &24--23                    & 1524 & 108.8 &   $0.0\pm0.3$ &  $-3.7\pm0.3$     & $12.2\pm0.5$  &   $17.0\pm0.6$  \\
& \multicolumn{2}{c}{Continuum}                   &--  & 108.5 &    $-0.05\pm0.5$ & $-0.8\pm0.1$ &  $12.5\pm0.2$ &   $11.3\pm0.4$  \\
& \multicolumn{2}{c}{Continuum}                   &--  & 189.9 &    (0.0)  & (0.0)                                  & $16.5\pm0.5$  & $14.4\pm0.3$\\
\hline
\end{tabular}
\label{tab:linemaps}
\end{table*}

\begin{figure}[!t]
\includegraphics[width=90mm]{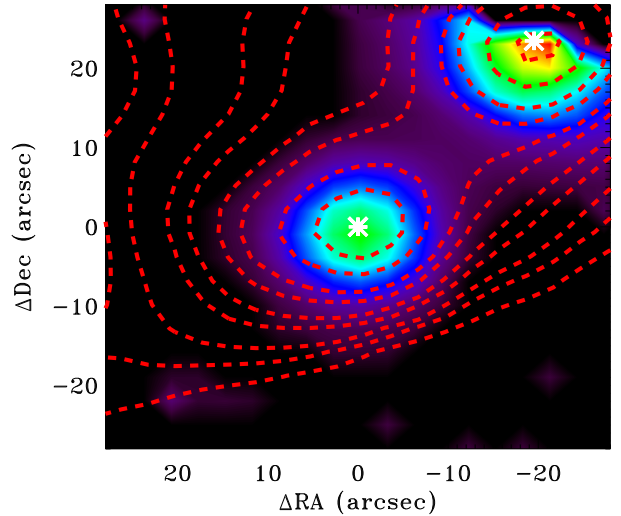}
\caption{The 63 $\mu$m continuum emission (colors) and the SCUBA 450 $\mu$m emission map (contours), with the location of IRAS 4A and IRAS 4B marked as white asterisks.  The PACS maps are shifted so that the centroid of the 190 $\mu$m continuum emission is located at the position of IRAS 4B obtained from sub-mm continuum interferometry.  The SCUBA map is shifted to the same position.}
\label{fig:iras4b_scuba.ps}
\end{figure}

All lines in the complete spectral scan are spectrally unresolved.  { The lines were initially fit with Gaussian profiles, with the central wavelength, line width, and amplitude and a first-order continuum as free parameters.  The instrumental line widths were then calculated from first-order fits to the wavelength-dependent spectral resolution, obtained from strong lines (Fig.~\ref{fig:specres.ps}).  Our final fluxes are obtained from fitting Gaussian profiles to each line, with the line width set by the calculated instrumental resolution at the given wavelength.}   For strong lines, the median centroid velocity is 30 \kms\ with a standard deviation of 24 \kms.  The absolute wavelength calibration is accurate to $\sim 50$ \kms\ and is limited by the spatial distribution of emission within each spaxel.

\begin{figure*}[!t]
\includegraphics[width=180mm]{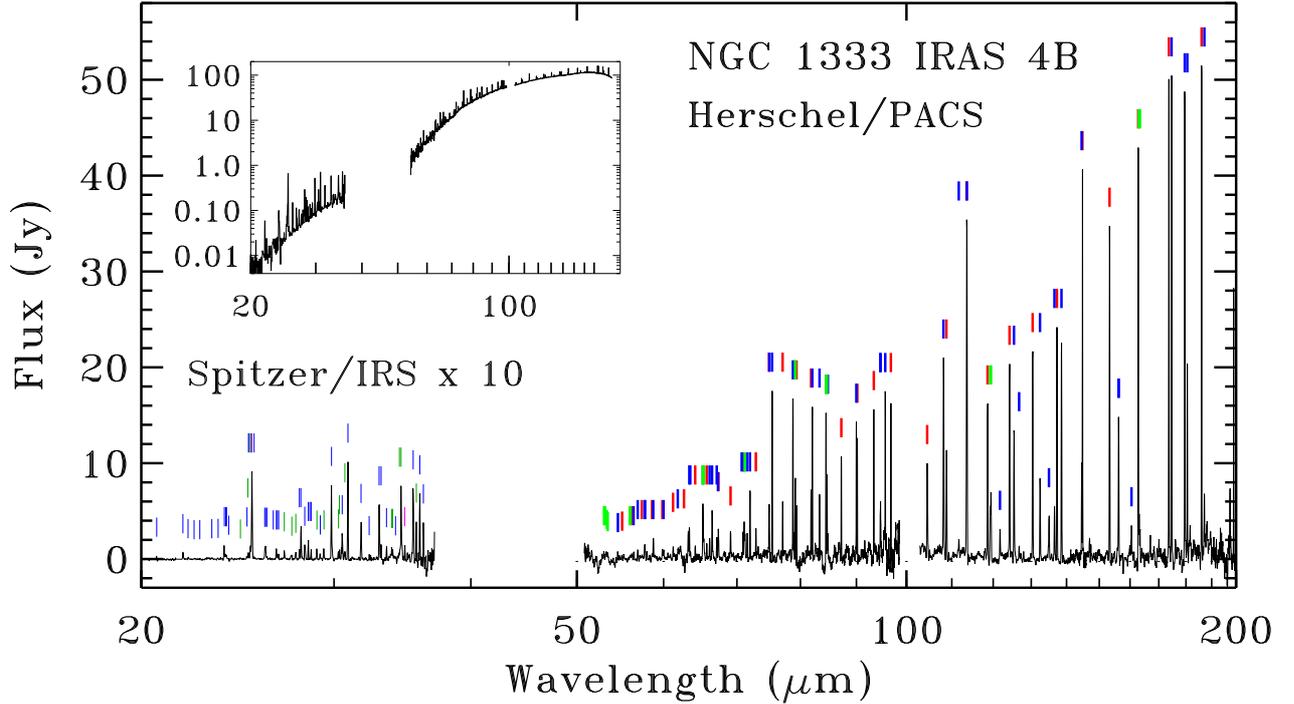}
\vspace{5mm}
\caption{The combined {\it Herschel}/PACS and {\it Spitzer}/IRS continuum-subtracted spectra of IRAS 4B, with bright emission in many H$_2$O (blue marks), CO (red marks), OH  (green marks), and atomic or ionized lines (purple).  The {\it Spitzer} spectrum is multiplied by a factor of 10 so that the lines are strong enough to be seen on the plot.  The inset shows the combined spectrum including the continuum.}
\label{fig:sed.ps}
\end{figure*}

\begin{figure}
\includegraphics[width=90mm]{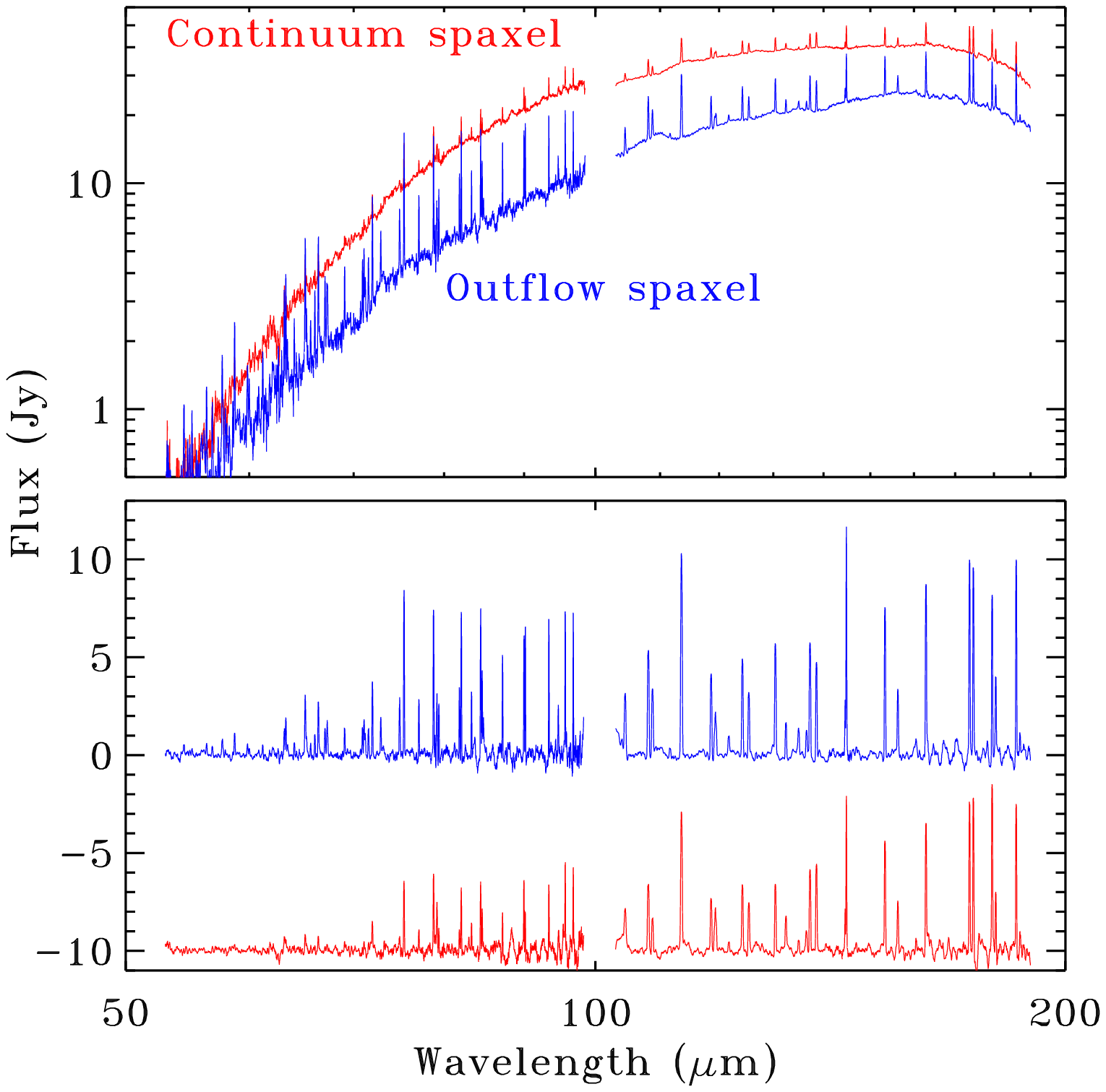}
\vspace{5mm}
\caption{{ Top:}  Spectra extracted separately from a spaxel centered on the sub-mm continuum (red) and a spaxel offset by $9\farcs4$ to the S and centered on the outflow position (blue).  The outflow position is dominated by line emission while the central object is dominated by continuum emission, indicating that the lines and continuum are spatially offset.  { Bottom:}  The same two spectra after continuum subtraction show that the on-source line emission gets much weaker to short wavelengths.}
\label{fig:twospax.ps}
\end{figure}

\subsection{{\it Herschel}/PACS Nyquist-sampled spectral imaging}

Nyquist-sampled spectral maps of narrow spectral regions at 54.5, 63.3, 108.5, and 190 $\mu$m were obtained in a total of 2.2 hr of integration time.  These maps were obtained from a $3\times 3$ raster scan with $3^{\prime\prime}$ steps, yielding spatial resolutions of $\sim 5^{\prime\prime}$, $5^{\prime\prime}$, $8^{\prime\prime}$, and $10^{\prime\prime}$, respectively.  Small grating steps were used to fully sample the spectral resolution.  The final spectrum is rebinned onto a wavelength grid with $\sim 4$ pixels per resolution element.

The spectral maps include CO, H$_2$O, and [O I] lines listed in Table \ref{tab:linemaps}.  In a $3\times 3^{\prime\prime}$ area, the typical rms is $\sim 0.02$ Jy per resolution element at 63 $\mu$m and $\sim 0.01$ Jy per resolution element at 108 $\mu$m.  This sensitivity level is better than that from the full spectral scan because of different integration areas and much longer integration times in each resolution element.

The data cubes from the Nyquist-sampled maps were reprojected onto a normal grid of right ascension and declination.  In the automated calibration the 190 $\mu$m continuum emission is offset from the object position$^1$, as measured from sub-mm interferometry \citep{Jorgensen2007}. 
Each  map is shifted in position by $3\farcs3$ E and $1\farcs5$ N so that the location of the 190 $\mu$m continuum from our observations matches the peak location of the sub-mm emission.  
After the shift, the 63 and 108 $\mu$m continuum emission from IRAS 4B and IRAS 4A (located in the NW edge of the map) are well aligned with the {\it Spitzer}-MIPS 70 $\mu$m emission.  No significant offset was measured in the staring observation of the full PACS SED, which was obtained with a new pointing one day after the Nyquist-sampled map.  Because the same spatial shift is applied to both the line and  far-IR continuum emission maps and because the shift moved both the line and continuum closer to the sub-mm continuum peak, the result that the line emission is spatially offset from the sub-mm continuum is robust to the pointing uncertainty.
\footnotetext[1]{The optical guider observations onboard {\it Herschel} are typically accurate to $\sim 1^{\prime\prime}$.  However, observations of nearby high-extinction regions, including NGC 1333 IRAS 4B, have few optical guide stars, which may introduce pointing offsets when the few guide stars are distributed asymmetrically in the field-of-view.}

\begin{figure*}[!t]
\includegraphics[width=60mm]{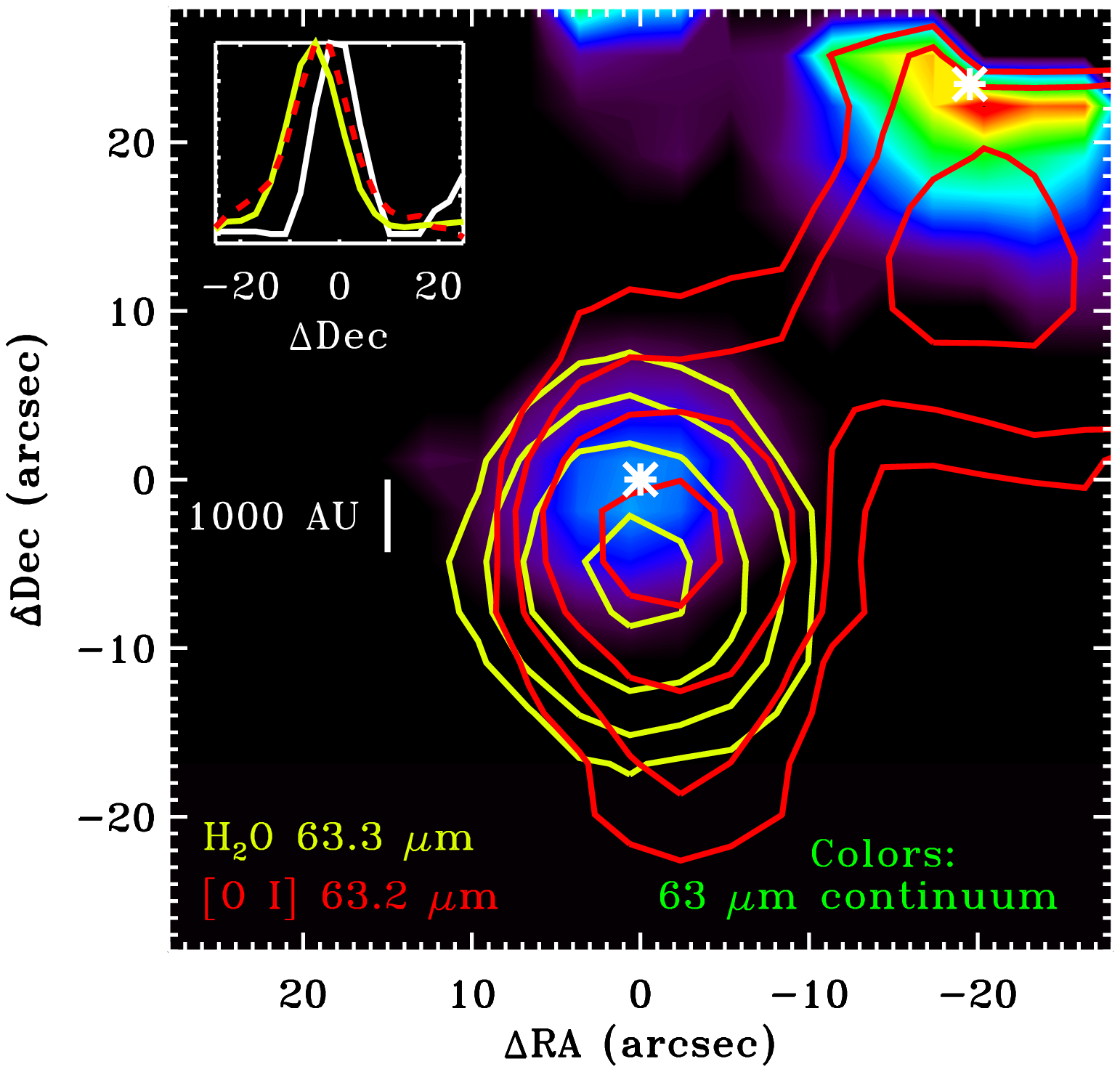}
\includegraphics[width=60mm]{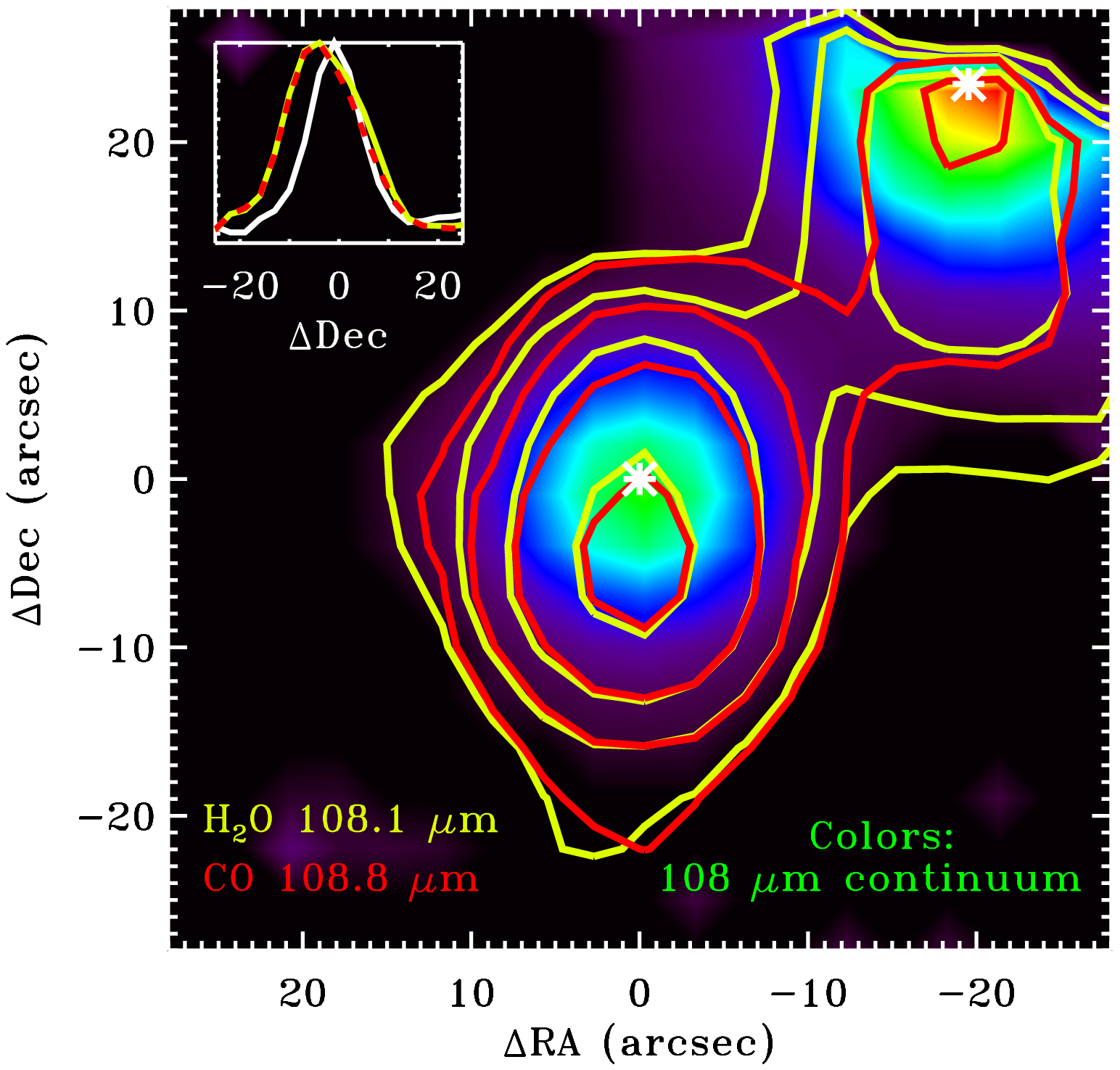}
\includegraphics[width=60mm]{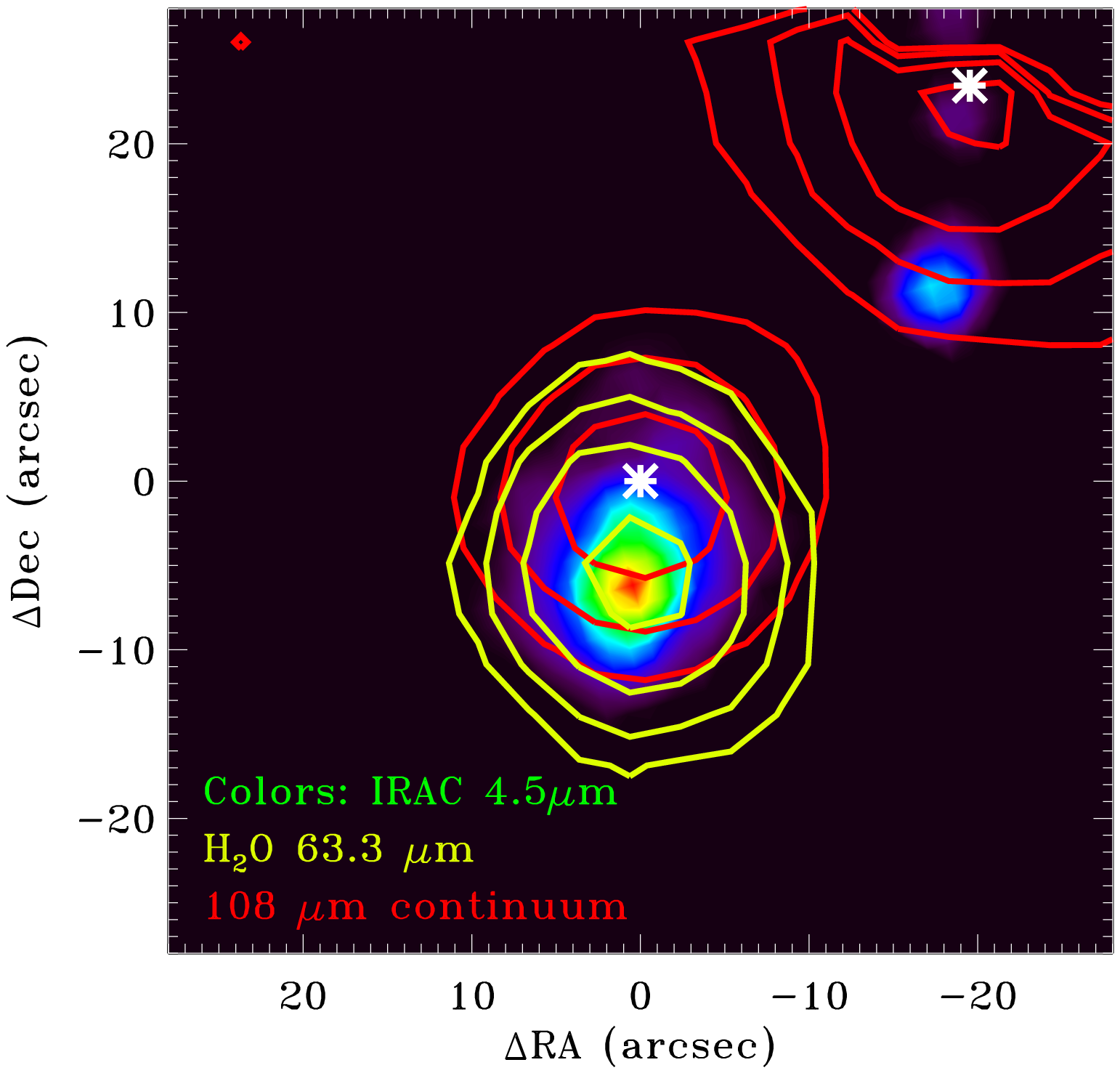}
\caption{{ Left and Middle:}  Contour maps of H$_2$O (yellow),  [O I] (red, left), and CO (red, middle) emission compared with continuum emission (color contours) at 63.3 $\mu$m (left) and 108.5 $\mu$m (middle).  The insets show the spatial extent of continuum and line emission in the different components along the N-S outflow axis. { Right:} a comparison of the location of emission in H$_2$O 63.32 $\mu$m (blue), 108.5 $\mu$m continuum (red), {\it Spitzer}-IRAC 4.5 $\mu$m photometry \citep[color contours;][]{Jorgensen2006,Gutermuth2008}.   All contours have levels of 0.1,0.2,0.4,0.8, and 1.6 times the peak flux near  IRAS 4B.  The asterisks show the sub-mm positions of IRAS 4B near the center of the field and IRAS 4A in the NW corner.  Some noise in these maps is suppressed at empty locations far from IRAS 4A and IRAS 4B.}
\label{fig:maps.ps}
\end{figure*}

The southern nod position in the map has spatially-extended [O I] emission in the southwest portion of the map.
The [O I] emission is therefore measured from only the northern nod position.  Inspection of {\it Spitzer}/MIPS 70~$\mu$m maps at the two nod positions does not indicate the presence of a strong, point-source continuum emission that could otherwise corrupt the continuum map for IRAS 4B.  { The 54 $\mu$m continuum map has low S/N and is not used.}

The two H$_2$O and [O I] lines at 63 $\mu$m have spectral widths of 110 \kms, which places an upper limit of 65 \kms\ on the intrinsic line width.  That the line widths are broader than the instrumental resolution of $\sim 3300$ is not significant because emission that is spatially extended in the cross-dispersion direction can broaden the spectral line profile, as with any other slit spectrograph.

\subsection{Spitzer/IRS Spectrum}

The {\it Spitzer}/IRS spectra of IRAS 4B were originally presented by \citet{Watson2007}.  We re-reduced the spectrum following the procedure described by \citet{Pontoppidan2010}.  The LH slit width is $11^{\prime\prime}$.  The flux extraction region of 5--10$^{\prime\prime}$ across the wavelength region was selected to optimize the final signal-to-noise in the limit of a point source \citep{Horne1986}.  The {\it Spitzer} and {\it Herschel} observations cover similar regions on the sky.  The relative flux calibration between {\it Spitzer} and {\it Herschel} spectra is likely uncertain by $\sim 30\%$ flux.

\section{RESULTS}

\subsection{Far-IR Spectrum of  IRAS 4B}

The far-IR PACS spectrum of IRAS 4B is the richest far-IR spectrum of a YSO to date.  A forest of high  signal-to-noise CO, H$_2$O, and OH lines that provide a full census of far-IR molecular emission that can be detected from low-mass YSOs (Fig.~4; see also Figs.~\ref{fig:specall1.ps}-\ref{fig:specall2.ps} in the Appendix).    Lines were discovered in the spectrum following a biased search for emission at the wavelengths for transitions of common species and an unbiased search for narrow features that peak above the noise level.

A total of 115 distinct emission lines are detected and identified from IRAS 4B (Table 3).  
All strong lines are identified.  Several tentative detections of weak lines are unidentified and discussed in Appendix B.
No H$_2^{18}$O or $^{13}$CO emission is detected in the PACS observations, with typical flux limits in the strongest expected lines of $\sim 0.03$ times the observed flux of the main isotopologue.  The [O I] 145.5 $\mu$m line is not detected, with a $2\sigma$ flux limit $1.2\times10^{-21}$ W cm$^{-2}$.  Lines of OH$^+$ and CH$^+$, HD 56 and 112 $\mu$m,  [N II] 121.8 and 205.2 $\mu$m, [C II] 157.7 $\mu$m, and [O III] 88.7 $\mu$m are also not detected.

Figure~\ref{fig:spaxmap.ps} shows a spectral map of the o-H$_2$O $6_{16}-5_{05}$ 82.03 $\mu$m ($E^\prime=643$ K) line emission overplotted on a {\it Spitzer}/IRAC 4.5 $\mu$m image of IRAS 4B (and IRAS 4A).   The line emission is located at the position of the near-IR emission in the blueshifted outflow lobe, south of the central source of IRAS 4B.
On the other hand, most of the continuum emission is located in the central spaxel, consistent with the location of the sub-mm continuum emission.  Figure \ref{fig:twospax.ps} demonstrates that the equivalent width of far-IR lines is much larger in the outflow spaxel than in the continuum spaxel, indicating a spatial offset between the line and continuum emission.  The line emission in the central spaxel mostly disappears at $<70~\mu$m.  All lines in the PACS spectrum are spatially offset from the continuum emission.  The line fluxes are measured based on the summation of these two bright spaxels and correction for spatial extent (see \S 2.1).

\subsection{PACS Mapping of H$_2$O Emission from IRAS 4B}

\begin{figure}[t]
\includegraphics[width=60mm,angle=90]{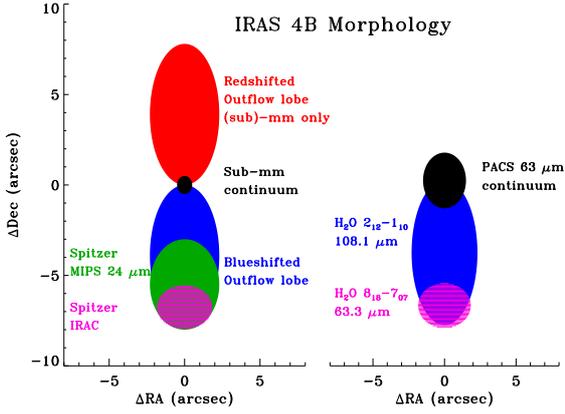}
\caption{A cartoon showing the location of different emission components from IRAS 4B, based in part on Fig.~\ref{fig:maps.ps}.}
\label{fig:cartoon.ps}
\end{figure}

Figure \ref{fig:maps.ps} and Table~\ref{tab:linemaps} compare the spatial distribution of the continuum flux with H$_2$O, CO, and [O I] line fluxes obtained from the Nyquist-sampled spectral maps.    The far-IR continuum emission is centered on-source, at the location of the sub-mm continuum, while line emission is centered to the south in the blueshifted outflow.   Figure~\ref{fig:cartoon.ps} shows a cartoon version of the approximate location of line and continuum emission and the morphology of IRAS 4B.  In the following analysis, we simplify the analysis by assuming that emission consists of two unresolved point sources, one at the outflow position and one at the sub-mm continuum peak, and are subsequently fit with 2D Gaussian profiles.   More complicated spatial distributions would be unresolved in our maps.  

The right panel of Figure~\ref{fig:maps.ps} demonstrates that the location of the warm H$_2$O coincides with the {\it Spitzer}/IRAC 4.5 $\mu$m imaging.  The two H$_2$O lines near 63.4 $\mu$m, o-H$_2$O $8_{18}-7_{07}$ and p-H$_2$O $8_{08}-7_{17}$, ($E^{\prime}=1070$~K) are centered at $5.2\pm0.2^{\prime\prime}$ from the 63 $\mu$m continuum and are spatially extended relative to the continuum emission (assumed to be unresolved for simplicity) by FWHM$=6.1\pm1.0^{\prime\prime}$. 
About 70\% of the emission is produced at the southern outflow position (Fig.~\ref{fig:crossflux.ps}).   The flux ratios for the two H$_2$O 63.4 $\mu$m lines are similar at both the on-source and off-source positions (Fig.~\ref{fig:specdepth.ps}).
In the 108.5 $\mu$m spectral map, both the o-H$_2$O $2_{21}-1_{10}$ ($E^{\prime}=194$~K) and CO 24--23 ($E^{\prime}=1524$ K) emission are centered $2.5\pm0.4^{\prime\prime}$ south of the 108.5 $\mu$m continuum emission and are spatially-extended in the outflow direction by $13.6\pm0.7^{\prime\prime}$, relative to the extent of the continuum emission.  The larger spatial extent and smaller offset in the 108 $\mu$m lines both indicate that the outflow component contributes $\sim40$\% of the measured line flux.  The spatial differences may be interpreted as differential extinction across the emission region, discussed in the next subsection.

The 54 $\mu$m maps are noisy because PACS has poor sensitivity at $<60$ $\mu$m.  The o-H$_2$O $5_{32}-5_{05}$ 54.507 $\mu$m ($E^{\prime}=732$~K) emission is offset by $5.9\pm0.4^{\prime\prime}$ south from the the peak of the sub-mm continuum emission.  The CO 49-48 53.9 $\mu$m emission ($E^{\prime}=6457$~K) is offset $2.9\pm0.4^{\prime\prime}$ south, between the peak of the sub-mm emission and the bright outflow location.  The highly-excited CO emission is produced in a different location than the highly-excited H$_2$O emission.

 The [O I] emission is offset by $3.7\pm0.3^{\prime\prime}$ at PA=$168^\circ$, just west of the outflow, and is spatially extended by $\sim 7\farcs1\pm1.0$.

\subsection{Extinction Estimates to the Central Source and Outflow Position}

The extinction to different physical structures within the IRAS 4B system depends on how much envelope material is present in our line of sight.  In this subsection, we discuss how these different extinctions affects the emission that is seen.  The extinction law used here is obtained from \citet{Weingartner2001} with a total-to-selective extinction parameter $R_V$=5.5, typical of dense regions in molecular clouds \citep[e.g.][]{Indebetouw2005,Chapman2009}.  Appendix D includes a discussion of how extinctions may affect the molecular excitation diagrams.

The strength of the near-IR emission in the outflow  \citep{Jorgensen2006} indicates that extinction must be low to at least some of the outflow position.  Spherical models of the dust continuum indicate that the central protostar is surrounded by $A_V=1000$ mag.~\citep{Jorgensen2002}, so any emission from the redshifted outflow lobe may suffer from as much as $A_V\sim 2000$ mag.~of extinction.  Any additional extinction would have likely introduced asymmetries in the H$_2$O line profiles that were presented in \citet{Kristensen2010}.  Depending on the wavelength and spatial location of the emission, the far-IR emission line fluxes may be severely affected by extinction.

The H$_2$O 54.5 and 63.4 $\mu$m lines have a different spatial distribution than the H$_2$O 108.1 $\mu$m line.  If we assume that the on-source and off-source emission both have similar physical conditions, then the ratio of the H$_2$O 63.4 to 108.1 $\mu$m line luminosities should not change with position.  In this scenario, the different locations for the detected flux is caused by differential extinction across the emission area.  An average extinction to the on-source H$_2$O component of $A_V\sim 700$ mag.~would reduce the fractional contributions from the on-source and off-source locations observed values.  This extinction may be the combination of a lightly-extincted region on the front side of the protostar and a heavily-extincted  region on the back side of the protostar.

An independent estimate of the extinction can also be made from the flux ratio of [O I] 63.18 to 145.5 $\mu$m lines, which is typically observed to be about 10 \citep{Giannini2001,Liseau2006}. 
 The undetected [O I] 145.5 $\mu$m line flux is less than 10\% of the $1.8\times10^{-20}$ W cm$^{-2}$ flux in the [O~I] 63.18 $\mu$m line. 
 If we conservatively assume that the true ratio is 30, then we estimate $A_V<200$ mag.~to the [O~I] emission region.  Some additional [O~I] emission could only be hidden behind a high enough extinction ($A_V\sim4000$ mag.) to attenuate emission in both the 63.18 and 145.5 $\mu$m lines.  Therefore, the effect of extinction on the [O I] luminosity is likely not too significant for the southern outflow lobe,  where [O~I] emission is seen.

\subsection{{\it Spitzer}/IRS Spectrum and Broadband Images of IRAS 4B}

The {\it Spitzer}/IRS spectrum of IRAS 4B includes emission in lines of highly-excited H$_2$O and OH, plus [S I] and [Si II].  Our line identification mostly agrees with that of \citet{Watson2007} for H$_2$O lines, with some modifications to account for the identifications of OH lines (Table~\ref{tab:spitzeroh} and Figure~\ref{fig:ohplot.ps}, see also Tappe et al.~2008).  The H$_2$O line identification was informed from line intensities predicted by RADEX modelling (see \S 4.1) of both the low-density case discussed here and the high density-case of \citet{Watson2007}.  All lines with significant detections are identified.

\begin{figure}
\includegraphics[width=90mm]{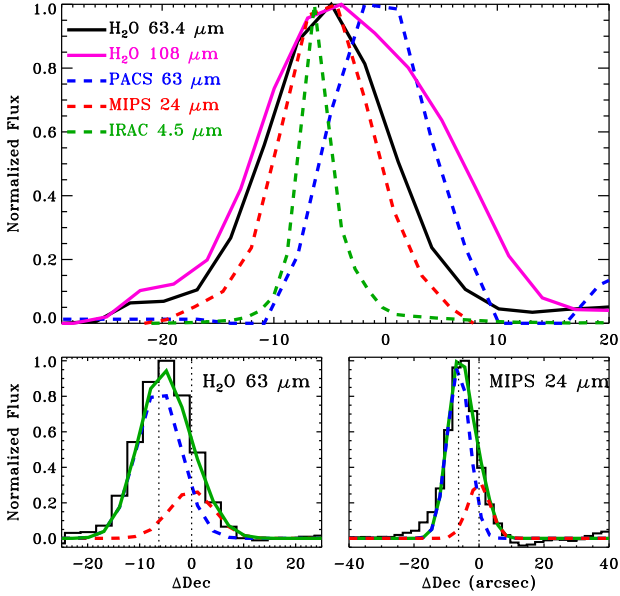}
\caption{{ Top:}  A cross cut of the flux at several wavelengths along the N-S outflow axis.  The 63 $\mu$m continuum is centered at the peak of the sub-mm continuum while the 4.5 $\mu$m and 24 $\mu$m photometry are centered in the outflow.  The H$_2$O 63.3 $\mu$m lines are produced primarily at the outflow location while the H$_2$O 108.1 $\mu$m line has similar on- and off-source contributions.  The (0,0) position is defined here by the peak of the sub-mm continuum emission from \citet{Jorgensen2007}.  { Bottom:}  The spatial cross cut of H$_2$O 63.3 $\mu$m and MIPS 24 $\mu$m emission, shown as the combination of two unresolved Gaussian profiles located at the on-source position (red dashed line) and the blueshifted outflow position (blue dashed line).}
\label{fig:crossflux.ps}
\end{figure}

{ 
\begin{table*}
\caption{OH lines detected in the {\it Spitzer}/IRS spectrum of IRAS 4B$^a$}
\begin{tabular}{cccccc}
\hline
Line ID & $E_{up}$ (K) & $\log A_{ul}$ (s$^{-1}$) & $\lambda_{vac}$ ($\mu$m) & Flux$^b$ & Err$^b$\\
\hline
$^2\Pi_{1/2}-^2\Pi_{3/2}~J=9/2-7/2^c$ & 875 & -1.40 & 24.62 &  1.6 & 0.2\\
 $^2\Pi_{3/2}~J=21/2^--19/2^+$    & 2905 & 1.29 & 27.39 & 6.0 & 0.4\\
 $^2\Pi_{3/2}~J=21/2^+-19/2^-$    & 2899 & 1.29 & 27.45 & 8.7 & 0.4\\
 $^2\Pi_{1/2}~J=19/2-17/2^c$       & 2957  & 1.28 & 27.67 & 8.3 & 0.2\\
 $^2\Pi_{1/2}-^2\Pi_{3/2}~J=7/2-5/2^c$    & 617 & -1.50 & 28.94 & 8.3 & 0.2\\
 $^2\Pi_{3/2}~J=19/2^+-17/2^-$      & 2381 & 1.16 & 30.28 & 11.9 & 0.3\\
 $^2\Pi_{3/2}~J=19/2^--17/2^+$       & 2375 & 1.15 & 30.35 & 13.1 & 0.3\\
 $^2\Pi_{1/2}~J=17/2^+-15/2^-$        & 2439 & 1.14 & 30.66 & 7.2 & 0.6\\
 $^2\Pi_{1/2}~J=17/2^--15/2^+$         & 2436 & 1.14 & 30.71 & 10.9 & 0.6\\
  $^2\Pi_{3/2}~J=17/2^--15/2^+$        & 1905 & 1.01 & 33.86 & 11.3 & 0.8\\
  $^2\Pi_{3/2}~J=17/2^+-15/2^-$        & 1901 & 1.00 & 33.95 & 8.4 & 0.8\\
 $^2\Pi_{1/2}~J=15/2-13/2^c$   & 1969 & 0.98 & 34.61 & 9.3 & 1.0 \\
\hline
\multicolumn{6}{l}{$^a$Listed OH lines are the sum of unresolved triplet hyperfine structure transitions.}\\
\multicolumn{6}{l}{$^b$$10^{-22}$ W cm$^{-2}$, with $1-\sigma$ error bars.}\\
\multicolumn{6}{l}{$^c$Includes two sets of unresolved triplets with different parities.}\\
\end{tabular}
\label{tab:spitzeroh}
\end{table*}
}

An analysis of the {\it Spitzer}/MIPS 24 $\mu$m image, with  sensitivity from 20--31 $\mu$m, helps us place limits on the amount of H$_2$O emission that could arise on-source.  Convolving the filter transmission curve with the {\it Spitzer}/IRS spectrum from \citet{Watson2007} indicates that $\sim 24$\% of the light in the MIPS 24 $\mu$m bandpass is in molecular emission (mostly H$_2$O), 17\% in the [S I] 25.24 $\mu$m line, and 59\% in the continuum.

 \citet{Jorgensen2010} found that the emission from IRAS 4B in the {\it Spitzer}/MIPS 24 $\mu$m images is offset from the peak emission of the sub-mm continuum \citep[see also][]{Choi2011}.   We measure that the emission is centered at $5\farcs2$ S and $0\farcs9$ E from the central source, consistent with the location of the outflow emission, and  is spatially extended by $\sim 5\farcs5$ in the north-south direction, along the outflow axis.  An additional component is present at the location of the sub-mm continuum peak.  The {\it Spitzer}/MIPS emission is assumed here to be a combination of emission from two unresolved sources, one at the sub-mm continuum peak and one at the position of the {\it Spitzer} IRAC 4.5 $\mu$m emission located 6\farcs2 S and 0\farcs4 E.  From fitting two dimensional Gaussian profiles to the image, the component at the blueshifted outflow lobe accounts for 78\% of the {\it Spitzer}/MIPS 24 $\mu$m emission and the sub-mm point source accounts for the remaining 22\% of the emission (see Fig.~\ref{fig:crossflux.ps} for the fit to the image collapsed onto the outflow direction).  Some additional {\it Spitzer}/MIPS emission is located at $20^{\prime\prime}$ S of the sub-mm continuum peak and is ignored here.

In the {\it Spitzer}/IRAC images of emission between 3.8-8 $\mu$m, the emission is located entirely at the outflow position.  In contrast, the 63 $\mu$m continuum emission is located mostly  on the central source at the sub-mm continuum position.  Much of the 20-31 $\mu$m continuum emission must be located at the outflow position, but some continuum emission could also be located on the central source.  Given the fraction of emission located on-source (22\%) and the relative contributions of molecular lines (24\%) and continuum to the {\it Spitzer}/MIPS photometry, the MIPS map could be consistent with an on-source location of H$_2$O emission only if the continuum emission is located entirely at the outflow position.

\begin{figure}
\includegraphics[width=90mm]{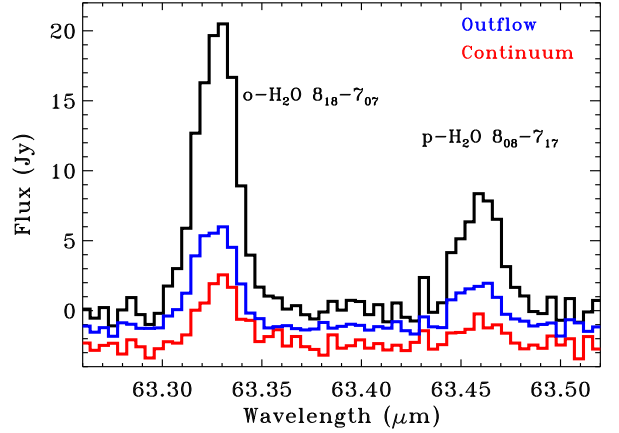}
\caption{The H$_2$O 63.4 $\mu$m spectral region extracted from the on-source position (red) and from the blueshifted outflow position (blue), and over the entire spectral map (black).  No significant differences are detected in the ratio of the two lines, which indicate that the two lines are optically thin at both locations.}
\label{fig:specdepth.ps}
\end{figure}

\section{EXCITATION OF MOLECULAR EMISSION FROM IRAS 4B}

The spatial distribution of the highly excited H$_2$O emission in the PACS observations places the bulk of the highly excited far-IR emission at the outflow position.  
In this sections, we analyze the excitation of the H$_2$O lines in detail to demonstrate that the highly excited H$_2$O emission in both the {\it Herschel}/PACS and {\it Spitzer}/IRS spectra can be explained with emission from a single isothermal, plane-parallel slab.  We subsequently analyze CO, OH, and [O I] emission from IRAS 4B.  Although the H$_2$O emission region is likely complicated and includes multiple spatial and excitation temperature components, our simplified approach is able to reproduce the highly excited H$_2$O lines.  The properties of this slab are the combination of the spatially offset outflow component and the on-source component.  
We lack sufficient spatial resolution throughout most of the spectrum to analyze the excitation of the two components separately. 

Figure~\ref{fig:excit.ps} shows excitation diagrams for H$_2$O, CO, and OH emission$^2$.  Without considering sub-thermal excitation, each molecule requires two temperature components to reproduce the measured fluxes.   For convenience, the two components for each molecule are called ``warm'' and ``cool'', however this terminology applies separately to each molecule$^3$.
The warmer component of CO may not be related to the warmer component of H$_2$O or OH.  For the temperature and density derived below, the two apparent excitation temperatures for OH and H$_2$O could even be produced by a single component, with the warm and cool regimes resulting from subthermal excitation.
Table~\ref{tab:cooling} describes the excitation temperatures, molecular column density, and luminosity for fits to these diagrams.
In the following subsections we discuss the excitation of the H$_2$O in detail, and briefly describe the excitation of CO, OH, and O.   RADEX models of H$_2$O are used to fit  only the higher excitation H$_2$O lines because the lower excitation lines are optically thick and difficult to use to infer physical conditions of the emitting gas.  The emitting area, temperature, and density derived from the fits to the observed H$_2$O lines are assumed to also apply to CO, OH, and O for simplicity.
\footnotetext[2]{Throughout the paper all logarithms are base 10, all units for column density are in cm$^{-2}$, and all units for density are cm$^{-3}$.  The units in excitation diagrams are in number of detected molecules rather than column density.}
\footnotetext[3]{The terminology ``warm'' and ``cool'' components, as defined here, corresponds to ``hot'' and ``warm'' gas, respectively, in other works, including \citet{Visser2011}, which considers colder gas, usually observed in the sub-mm, than the gas studied here.}

\begin{figure}[!t]
\includegraphics[width=90mm]{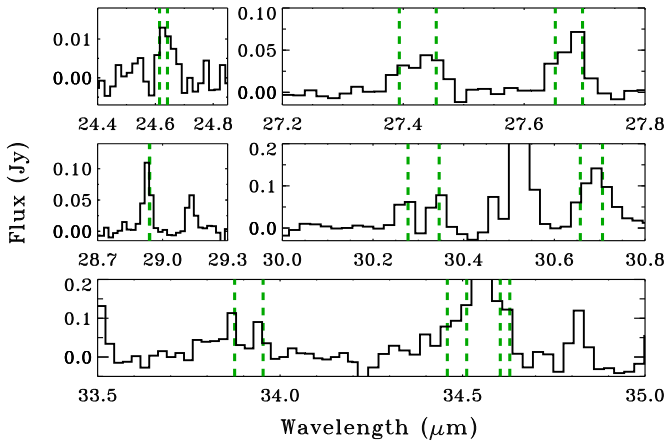}
\vspace{5mm}
\caption{OH emission lines (vertical dashed lines) in the {\it Spitzer}/IRS spectrum of IRAS 4B.  The lines expected to be strongest are either detected (Table~\ref{tab:spitzeroh}) or are blended with lines of other species.}
\label{fig:ohplot.ps}
\end{figure}

\subsection{RADEX Models of H$_2$O Emission}

\begin{figure*}[!t]
\includegraphics[width=90mm]{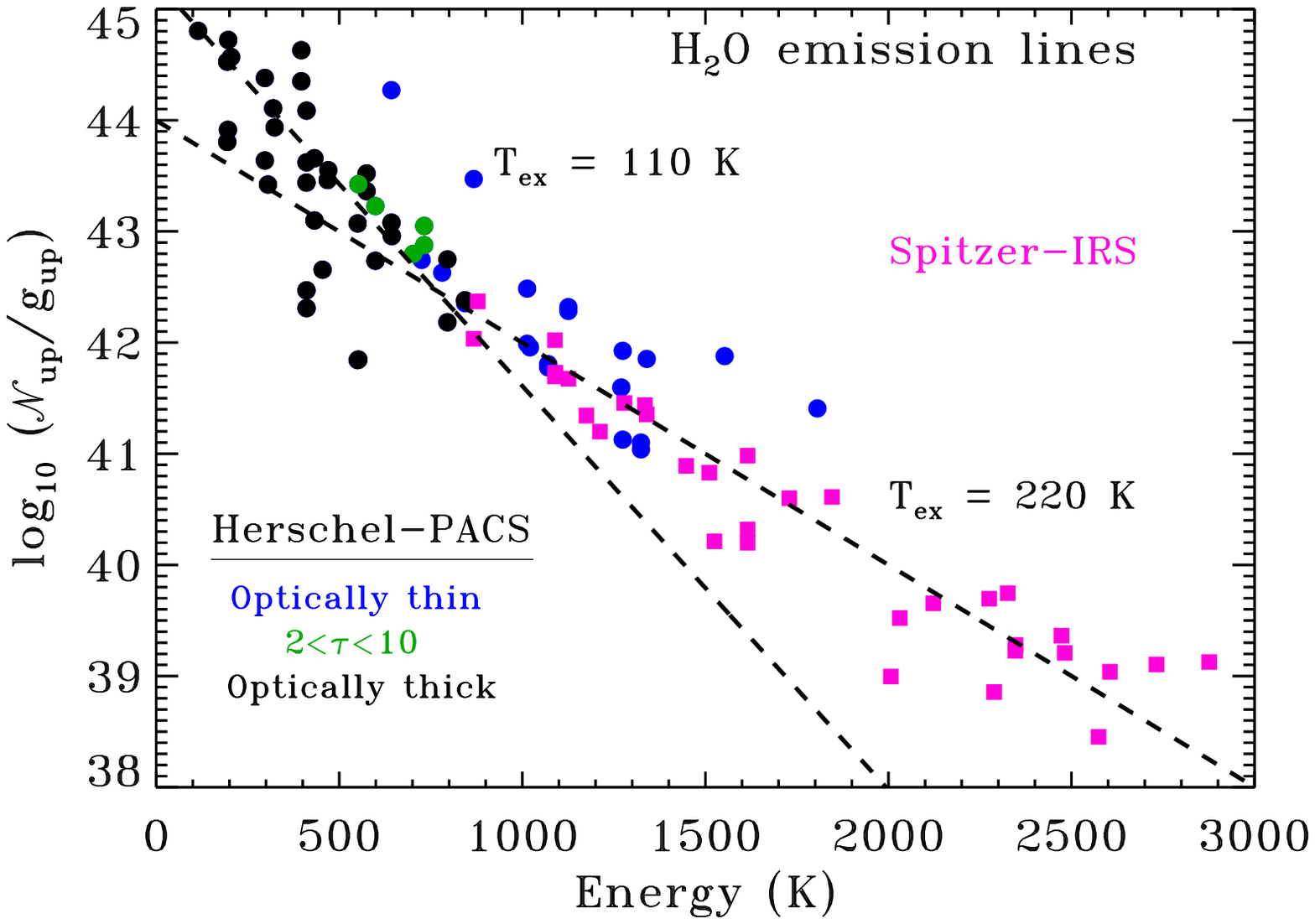}
\includegraphics[width=90mm]{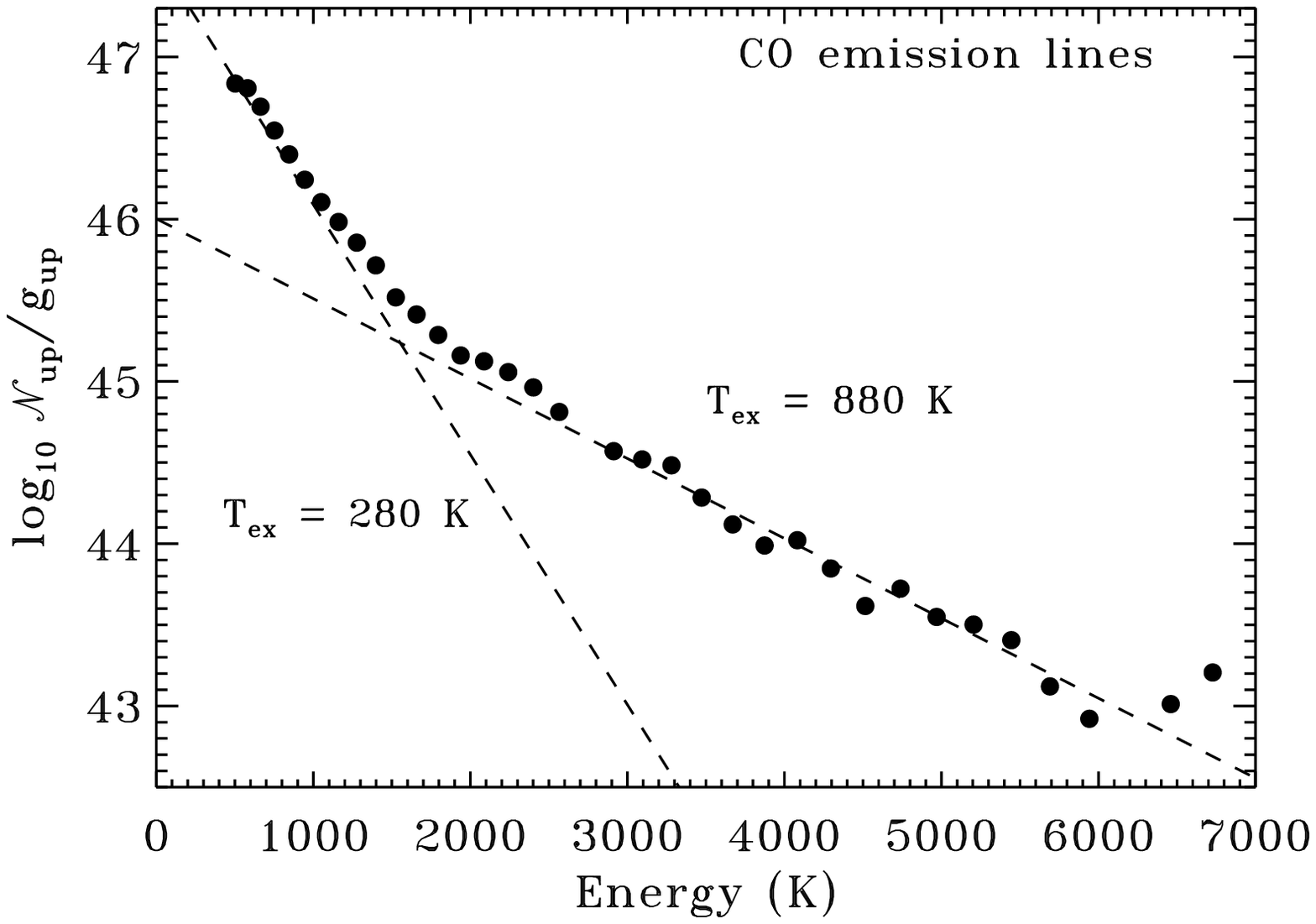}
\includegraphics[width=90mm]{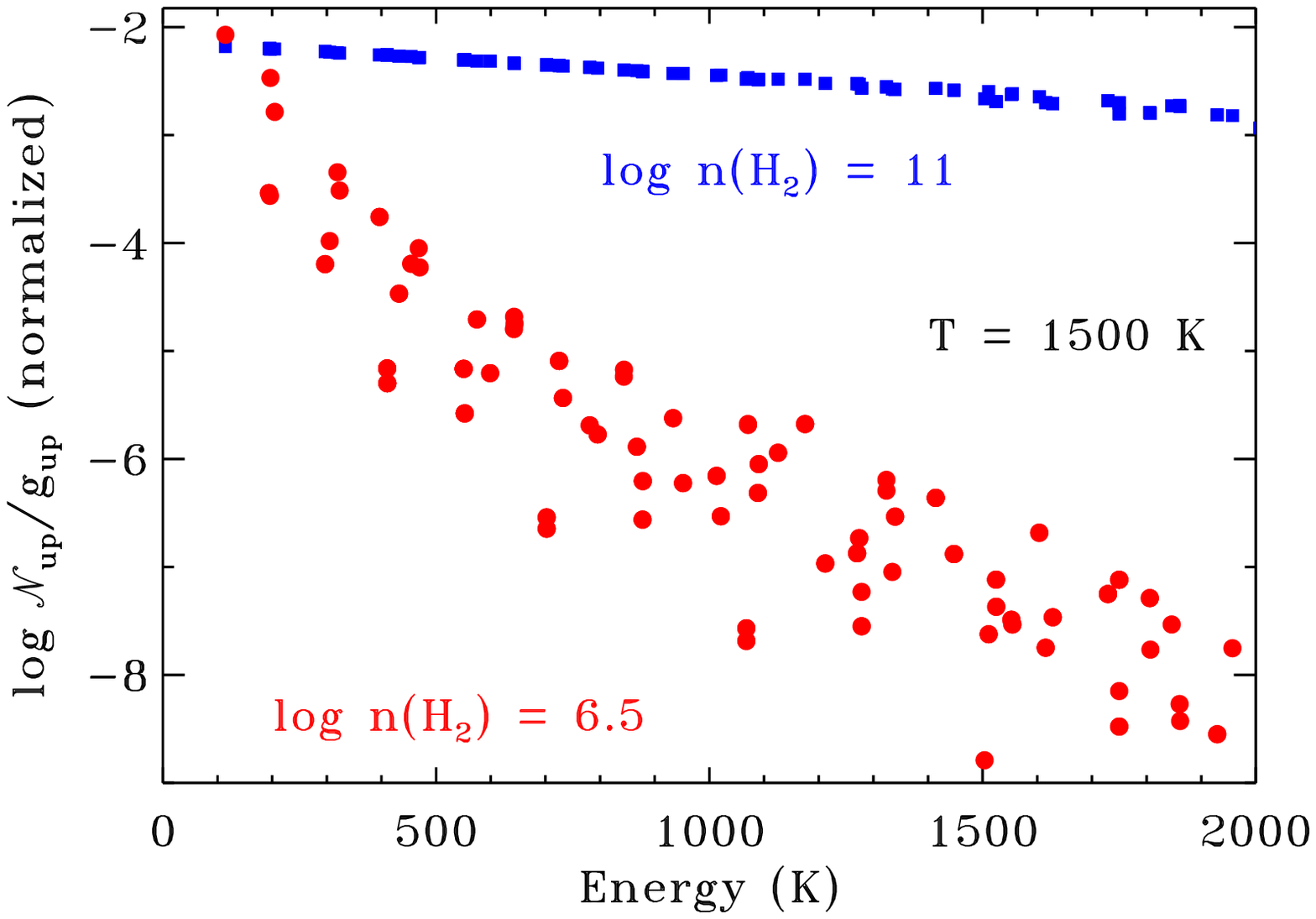}
\includegraphics[width=90mm]{fg11d.ps}
\caption{Excitation diagrams, in units of total number of detected molecules $\mathcal{N}$ divided by degeneracy $g$, for H$_2$O (upper left), CO (upper right), and OH (lower right) emission lines detected with {\it Herschel}-PACS (circles) and {\it Spitzer}-IRS (purple squares).  For the H$_2$O excitation diagram, PACS data are subdivided into lines that are optically-thin (blue), moderately optically-thick (green), and optically-thick (black), for the H12 model.  Most of the {\it Spitzer} lines are optically-thin. Lower left:  H$_2$O excitation diagrams obtained from RADEX models for optically thin models with $T=1500$ K at high (blue) and low (red) density.  At low density, subthermal excitation leads to cooler measured excitation temperatures and significant scatter in the level populations.}
\label{fig:excit.ps}
\end{figure*}

The H$_2$O line emission extends to high energy levels, with level populations indicating an excitation temperature of 220 K but with significant scatter.   Among the highly excited lines, the H$_2$O excitation diagram does not show any break at high energies that would indicate multiple excitation components. The highly-excited H$_2$O levels have high critical densities ($\sim 10^{11}$ cm$^{-3}$). 
The bottom left panel in Fig.~\ref{fig:excit.ps} demonstrates that a large amount of scatter in excitation diagrams may be explained by sub-thermal excitation.  The excitation temperature could therefore be the kinetic temperature of dense ($>10^{11}$ cm$^{-3}$) gas or could result from subthermal excitation of warmer gas with lower density.  Line opacities also increase the scatter in observed fluxes.

We calculate synthetic H$_2$O spectra from RADEX$^4$ models of a plane-parallel slab \citep{vanderTak2007} characterized by a single temperature $T$, density $n$(H$_2$), and H$_2$O column density $N$(H$_2$O) with an emitting surface area $A$.  RADEX is a radiative transfer code that simultaneously calculates non-LTE level populations and line optical depths for a plane-parallel slab to produce line fluxes.   A large grid was calculated using molecular data obtained from LAMDA \citep{Schoier2005,Faure2007}.  Since this molecular data file lacks the most highly excited lines detected with {\it Spitzer}, individual RADEX models were calculated at specific gridpoints using a much larger and more complete database with energy levels obtained from \citet{Tennyson2001}, radiative rates from the HITRAN database \citep{Rothman2009}, and collisional rates with H$_2$ from \citet{Faure2008}.

\begin{figure*}[!t]
\includegraphics[width=90mm]{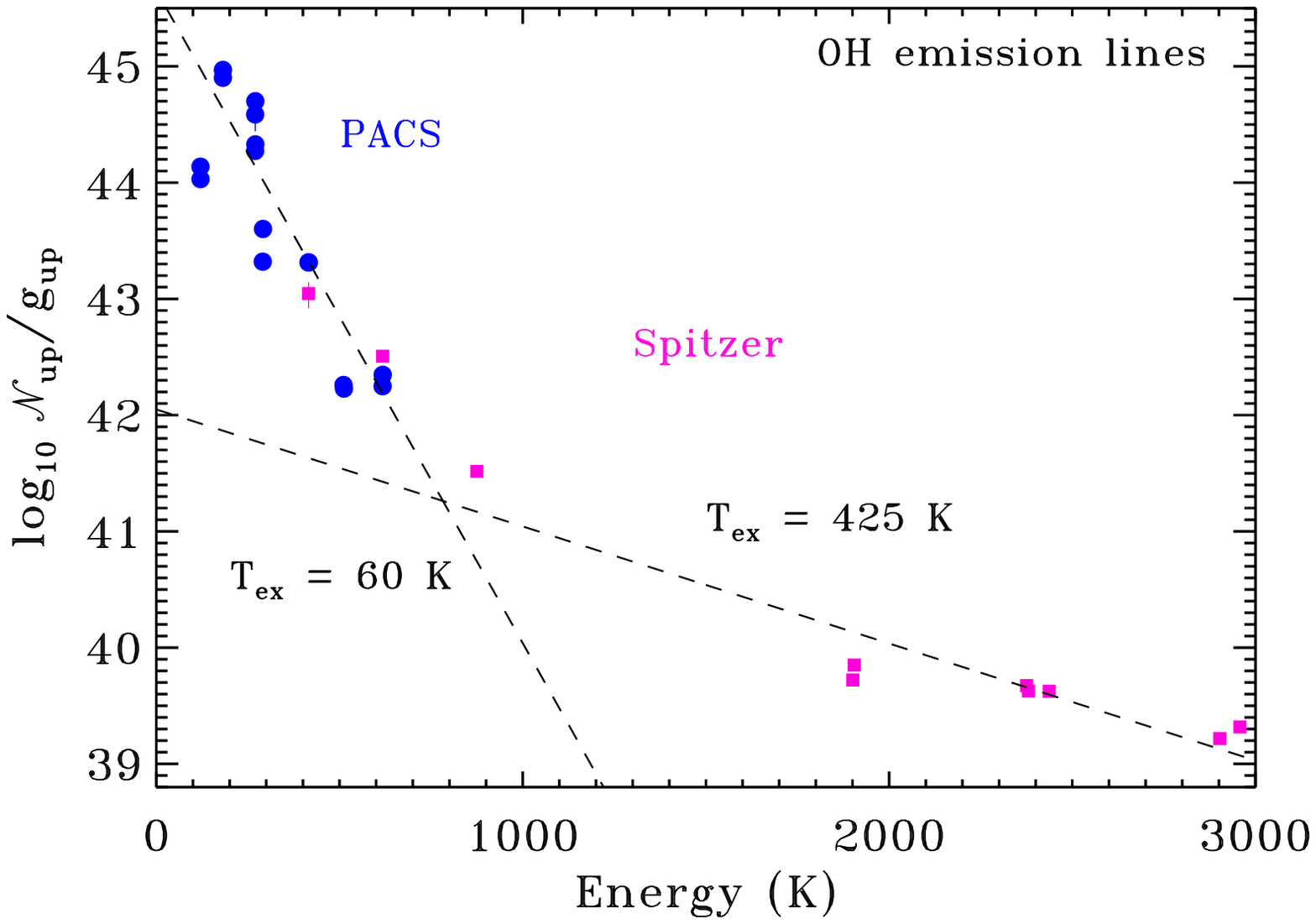}
\includegraphics[width=90mm]{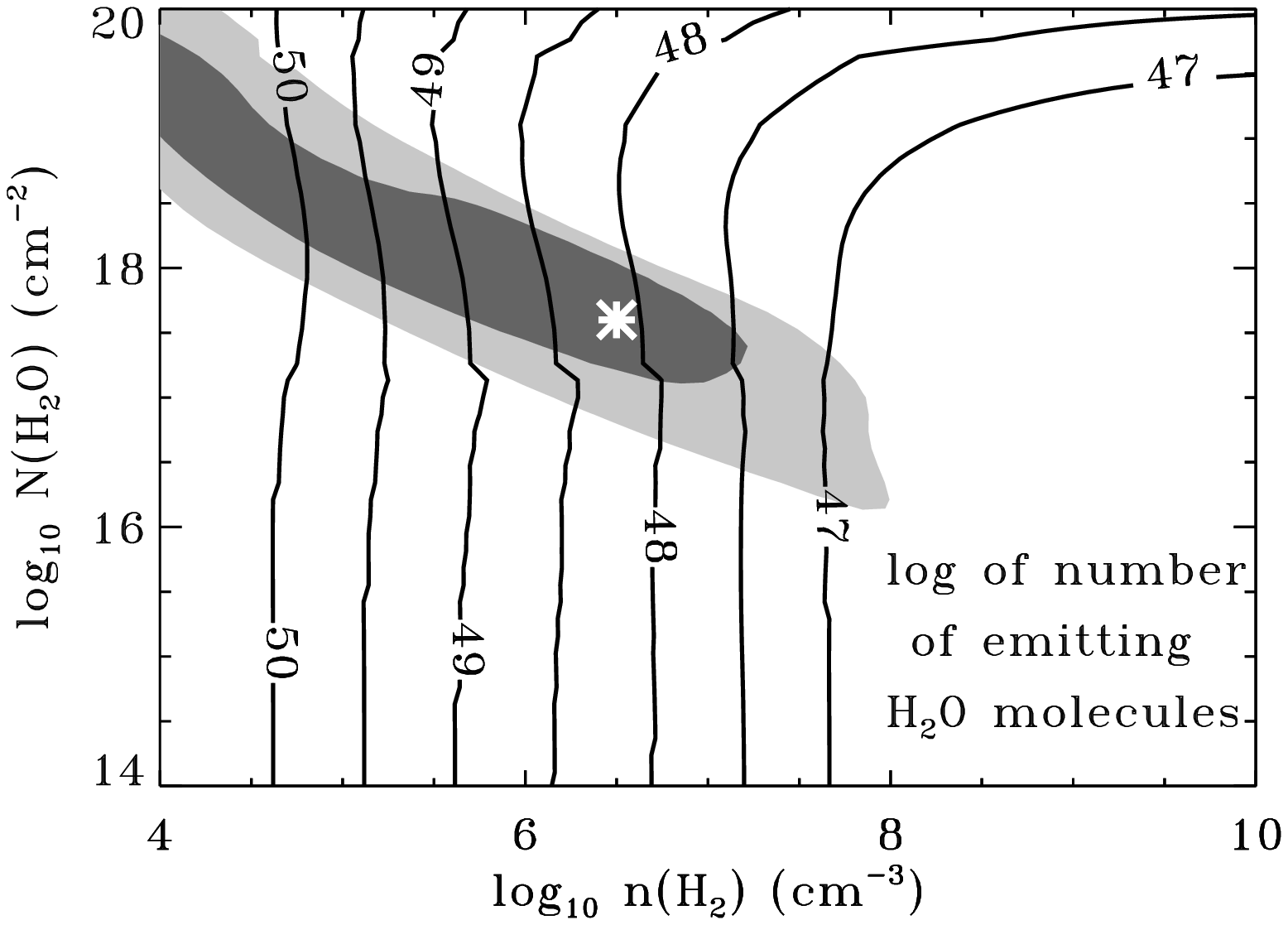}
\includegraphics[width=90mm]{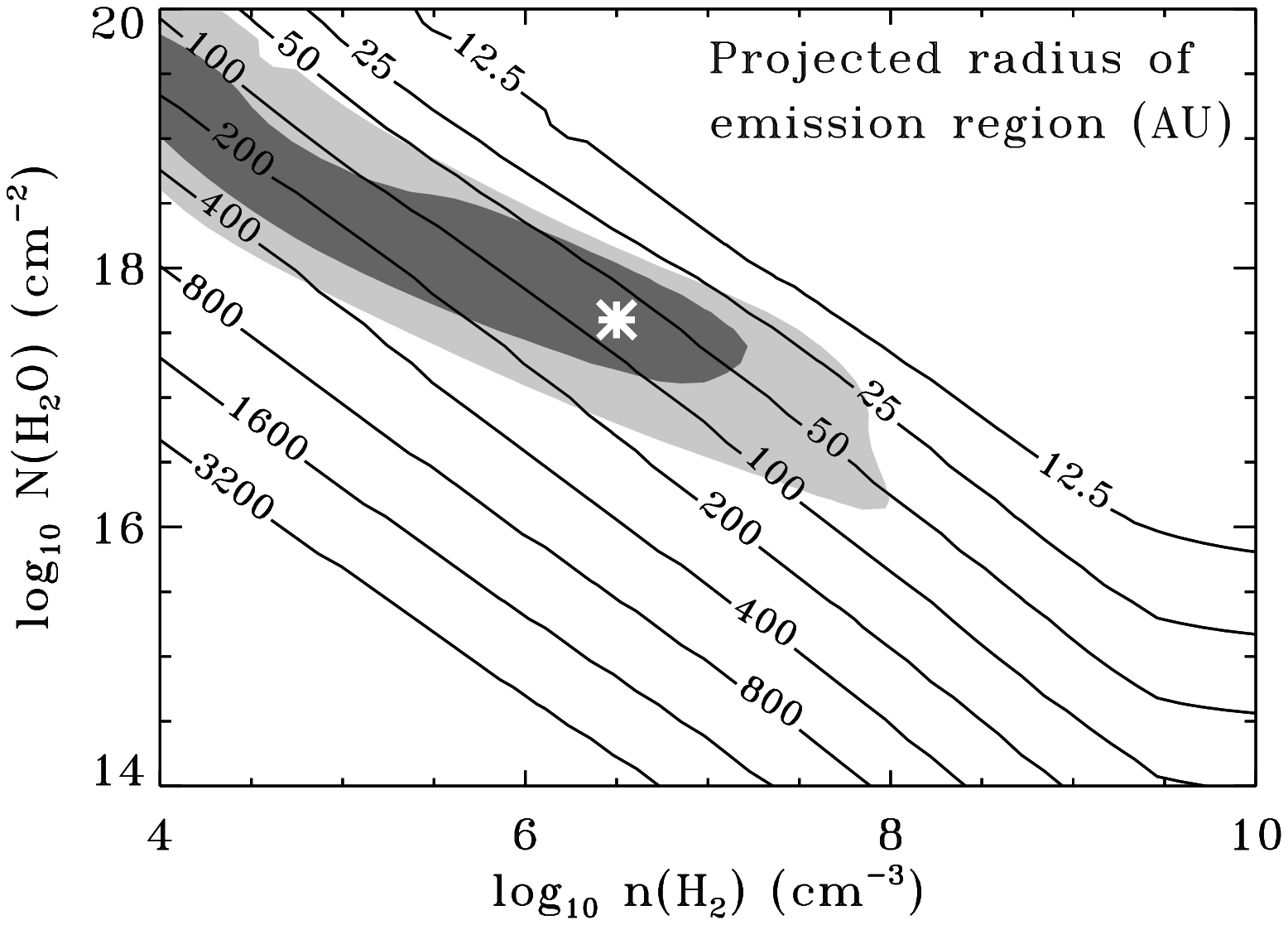}
\includegraphics[width=90mm]{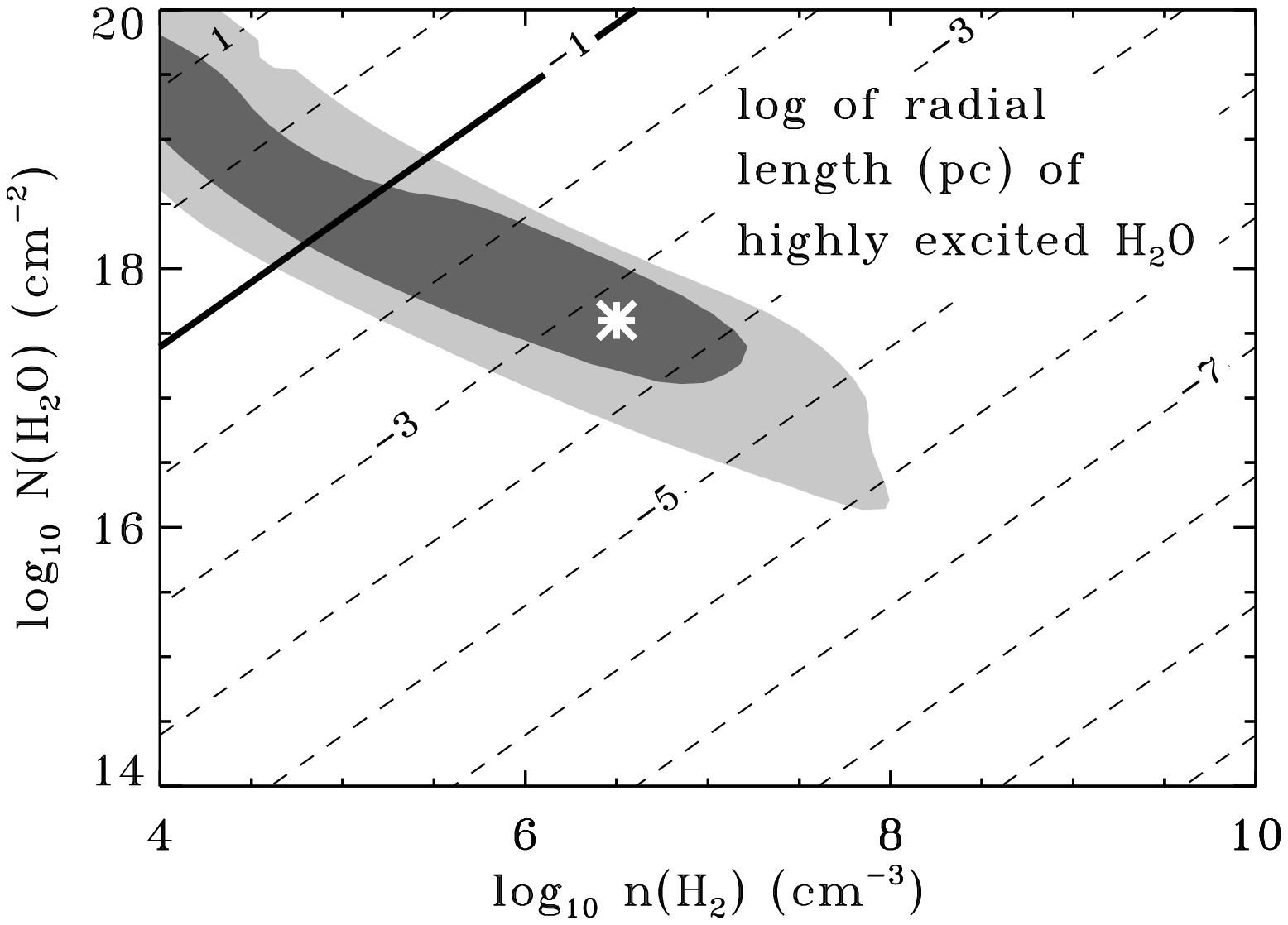}
\caption{{Upper left:}  Contours of $\chi^2$ versus temperature and H$_2$ density for several different H$_2$O column densities (the different colors), with the minimum reduced $\chi^2\sim4$.  The contours show where $\chi_{red}^2/4=1.2$ (solid lines) and $1.5$ (dashed lines), which roughly indicate the acceptable parameter space.  The contours shown here are calculated to fits of lines with $E^\prime>400$ K.  The asterisks show the parameters of the sub-thermal model presented here (H12) and the high density envelope-disk accretion shock model (W07).  Upper right and lower panels:  contours of total number of H$_2$O molecules (upper right), radius (AU) of a circular emission area for H$_2$O emission (lower left), and length scale (log pc) of H$_2$O emission (lower right) compared with the best fit contours of $n$(H$_2$) and $N$(H$_2$O) for $T=1500$ K.  The asterisks indicate the parameters adopted for this paper.  In the lower right panel, the solid line at 0.1 pc shows the conservative upper limit to the length scale of the emission.  The parameters adopted for this paper are located in the lower right of the acceptable parameter space so that the length scale for the emission is small.}
\label{fig:h2oexcit.ps}
\end{figure*}

The RADEX models are calculated to obtain a rough idea of the physical properties of the emitting gas.  Radiative pumping is not included in the model {but is likely important, especially at low densities.   Including radiative pumping would require detailed physical and chemical modeling of the envelope and is beyond the scope of this work.}  The line profile is assumed to be a Gaussian profile with a FWHM of 25 \kms,  based on the FWHM of low-excitation H$_2$O lines observed with HIFI \citep{Kristensen2010}.$^5$
The extinction to the warm H$_2$O gas is highly uncertain and is mostly ignored (see \S 3.3 and Appendix D for a discussion of extinction estimates and their implications).  The ortho-to-para ratio is assumed to be $3$, based on flux ratios of optically-thin lines that range from 2.8--3.5.
\footnotetext[4]{http://www.strw.leidenuniv.nl/$\sim$moldata/radex.html}
\footnotetext[5]{The HIFI H$_2$O lines are much more optically-thick and likely have larger emitting areas than most H$_2$O lines in the PACS spectrum, which may lead to differences in line profile.}

A $\chi^2$ fit (upper left panel of Fig.~\ref{fig:h2oexcit.ps}) to the measured fluxes with PACS and IRS lines with upper energy level above 400 K (the warm component) yields acceptable solutions with high temperature ($T>1000$ K) and H$_2$ densities of $\log n<7.5$.  Appendix C provides a detailed description of the line ratios and the non-detections of H$_2^{18}$O emission that constrain the best-fit parameters.
The size of the emission region further limits the range of acceptable parameter space (Fig.~\ref{fig:h2oexcit.ps}).  The emitting surface area $A$ is equivalent to the area of a circle with radius from 25--500 AU, which is reasonably close  to the projected length of the outflow on the sky ($\sim 1000$ AU or 0.005 pc).  If we assume that $N$(H$_2$O)$<10^{-4}$ $N$(H$_2$), then the given column density and H$_2$ density yields the length scales (depth along our line of sight) for the outflow that range from $>10^{-4}$ pc (for the acceptable model with the highest H$_2$ density and lowest H$_2$O column density) to $>130$ pc (for the model with the lowest H$_2$ density and highest H$_2$O column density).  Given the projected size of the outflow of 0.005 pc,  a length scale greater than 0.1 pc is uncomfortably large and rules out solutions with $\log n$(H$_2$)$<5$.  The lack of obvious vibrational excitation in spectra at 6~$\mu$m \citep{Maret2009,Arnold2011} limits the kinetic temperature to $\lesssim2000$ K.  The number of H$_2$O molecules scales with density.

\begin{figure*}
\includegraphics[width=180mm]{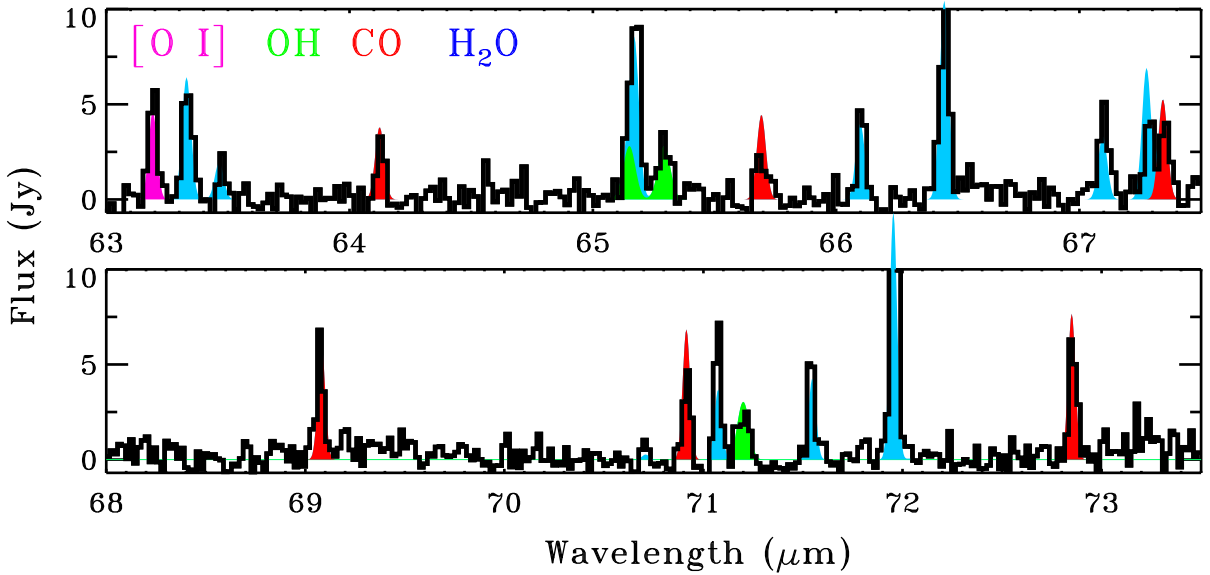}
\caption{Segments of the PACS spectrum (solid black line).  The H$_2$O line fluxes (shaded in blue) are obtained from RADEX, CO line fluxes (red) from the thermal distribution in Fig.~11, and OH (green) and [O I] line fluxes (purple) from Gaussian fits.}
\label{fig:segments.ps}
\end{figure*}

Combining these analyses, we adopt the parameters $T=1500^{+2000}_{-600}$ K, $\log n=6.5\pm1.5$, and $\log N$(H$_2$O)$=17.6\pm1.2$ over an emitting area equivalent to a circle with radius $25-300$ AU and length scale $0.003$ pc.  Figure~\ref{fig:segments.ps} shows that the PACS H$_2$O spectrum is well fit with model fluxes obtained with these parameters.  

The temperature and column density are both inversely correlated with the density, so the acceptable parameter space is tighter than implied by the large error bars.  The choice of line width scales the optical depth.  A broader line width would require the same factor increase in column density $N$(H$_2$O) and decrease in total emitting surface area.   The widths of the far-IR lines may differ from the optically thick low-excitation H$_2$O lines analyzed by \citet{Kristensen2010}.  The far-IR lines are primarily seen from the offset outflow location, while the longer-wavelength lines are dominated by on-source emission and include both red- and blue-shifted outflow lobes.  Very broad lines ($>65$ \kms) are ruled out from the widths of lines in the Nyquist sampled spectral maps.  

\subsection{Comparing H$_2$O spectra for warm, subthermal excitation and cool, thermalized gas}

\begin{figure}[!t]
\includegraphics[width=90mm]{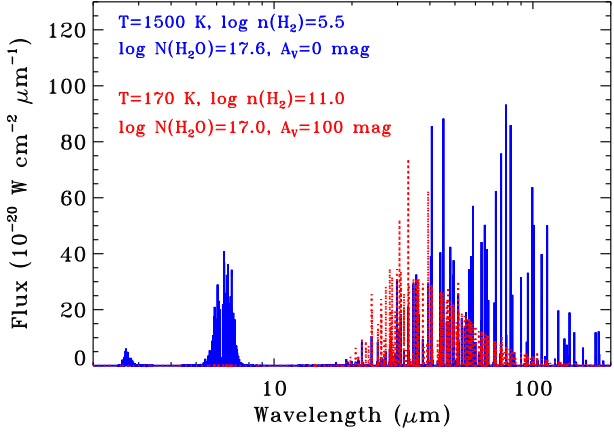}
\caption{The synthetic spectra for the H12 (blue) and W07 (red) models at a spectral resolution $R=1500$.  Because the W07 models are optically-thick in many highly-excited levels, the strongest H$_2$O lines peak in the mid-IR rather than the far-IR.  The H12 model predicts some rovibrational emission at 6 $\mu$m because of the high temperature, with a strength that could further constrain the temperature and density of the highly excited H$_2$O.}
\label{fig:synspec_all.ps}
\end{figure}

The high temperature, low density solution presented here (hereafter H12, with properties obtained from the $\chi^2$ fit listed above) produces sub-thermal excitation of H$_2$O, in contrast to the high-density (thermalized, with $\log n\sim11$, $\log N$(H$_2$O)$=17.0$, and $T=170$ K), low-temperature solution from \citet[][hereafter W07]{Watson2007}.  
The W07 slab has many optically-thick lines, which leads to a spectrum where the mid-IR H$_2$O lines observed with {\it Spitzer} are brighter than those in the PACS wavelength range  (Fig.~\ref{fig:synspec_all.ps}).  In contrast, many of the far-IR lines in the H12 parameters are optically-thin, so that most H$_2$O emission escapes in the PACS wavelength range.
As a consequence, in principle the far-IR H$_2$O emission could trace a high temperature, low density region (H12) while the mid-IR H$_2$O emission traces a high density, low temperature component (W07).

The W07 model produces optically-thick emission in the o-H$_2$O $8_{18}-7_{07}$ and p-H$_2$O $8_{08}-7_{17}$ lines at 63.4 $\mu$m, with a flux in the p-H$_2$O $8_{08}-7_{17}$ line similar to the observed flux.  However, the two lines are observed in an optically-thin flux ratio both on-source and at the blueshifted outflow lobe. 
Therefore, the W07 model cannot explain the PACS H$_2$O emission located at the outflow position.  If the extinction in W07 is reduced to $A_V\sim0$ mag, then the two 63.4 $\mu$m lines both become two times weaker relative to the mid-IR H$_2$O lines.

A comparison between the {\it Spitzer}/IRS spectrum and the synthetic H$_2$O spectrum (see Fig.~\ref{fig:comparison.ps} and a further discussion of line ratios in Appendix C) shows that both the W07 and H12 models could explain the mid-IR H$_2$O emission alone.
Both models are able to accurately reproduce the emission in most detected lines, with a few notable exceptions.  The p-H$_2$O $8_{35}-7_{26}$ 28.9 $\mu$m line flux is well reproduced in W07 but not H12.  However, the wavelength of this line is more consistent with an OH line than with the H$_2$O line.  The OH rotational diagram (Fig.~\ref{fig:excit.ps}) shows that the line flux is also consistent with fluxes in other OH lines with similar excitations.  The inability of H12 to reproduce this line flux with an H$_2$O model is therefore not significant.  An OH line at 24.6 $\mu$m was also misidentified as o-H$_2$O $8_{63}-8_{36}$ despite neither W07 nor H12 being able to produce flux in the H$_2$O line.  Because the synthetic fluxes of these lines are faint, the misidentification of this emission as H$_2$O can help to drive a best fit physical parameters to an optically thick solution.

In models with high $N$(H$_2$O) and $n($H$_2$), including W07, the line blend of o-H$_2$O $6_{52}-5_{23}$ and $5_{50}-4_{23}$ at 22.4 $\mu$m and the p-H$_2$O $6_{42}-5_{15}$ 23.2 $\mu$m line are predicted to be strong but are not detected.   Both W07 and H12 overpredict the flux in the H$_2$O 21.15 $\mu$m line.  The H12 model does not significantly overpredict any other line in the IRS spectrum, even if the best fit H12 fluxes are scaled to the level of the strongest IRS lines rather than to the far-IR PACS lines.  

This analysis demonstrates that a high temperature, low density model (H12) of the H$_2$O emitting region can reasonably reproduce both the PACS and IRS spectra.  On the other hand, a low temperature, high density model (W07) cannot reproduce the PACS spectrum, is inconsistent with the spatial distribution of the H$_2$O 63.4 $\mu$m line emission, and overpredicts the emission in several lines in the IRS spectrum.

\subsection{CO Excitation}

In the CO excitation diagram, a cool ($275\pm30$ K) component dominates mid-$J$ lines and a warm ($880\pm100$ K) component dominates high-$J$ lines.  The uncertainty in temperature includes the choice of energy levels to fit for the high and cool components.  The two excitation components could relate to regions with different kinetic temperatures or with different densities.  RADEX models of CO were run using molecular data obtained from LAMDA \citep{Schoier2005,Yang2010} with an extrapolation of collision rates up to $J=80$ by \citep{Neufeld2012}.  

The detection of high-$J$ CO lines with an excitation temperature of $\sim 880$ K requires $\log n$(H$_2$)$>6$ for reasonable kinetic temperatures ($<4000$ K).  For $T=1500$ K, $\log n$(H$_2$)$=6.5$, and an emitting area with radius 100 AU, the physical parameters adopted to explain the water emission, produces an excitation temperature of 950 K for lines with $J=30-45$.  This temperature is sensitive to the density, with $\log n$(H$_2$)$=6.0$ leading to an excitation temperature of 640 K, The total number of CO molecules, $\mathcal{N}$, for $\log n$(H$_2$)$=6.5$ is $\log \mathcal{N}=48.4$.

In principle the two temperature components could relate to a single region with high temperature ($\sim 4000$ K) and $\log n$(H$_2$)$<4$ \citep{Neufeld2012}.  In this case, the CO emission would be physically unrelated to the highly excited H$_2$O emission, and the CO abundance in the highly excited H$_2$O emission region would be much smaller than that measured here.


\subsection{OH Excitation}

The OH excitation diagram shows cool ($60\pm15$ K) and warm ($425\pm100$ K) components.  RADEX models of the low excitation levels of OH ($E_{up}<1000$ K) were run, using collisional rate coefficients from \citep{Offer1994} and energy levels and Einstein A values from \citep{Pickett1998}.   Molecular data for higher excitation levels were obtained from HITRAN \citep{Rothman2009}.  A $1500$ K gas with $\log$n(H$_2$)$=6.5$ and emitting area of radius 100 AU roughly reproduces the emission in the cool component OH emission, with $\log$N(OH)$=17.3$.  Whether these parameters could also reproduce the highly excited OH emission is not clear.

The RADEX model fluxes are also somewhat discrepant with the observed fluxes, The 24.6 $\mu$m line flux is much lower than predicted.  In addition, all detected OH doublets have similar line fluxes but the RADEX model predicts different fluxes in several transitions.  A lower opacity, caused either by broader lines or a lower column density and larger emitting area, would alleviate some of these discrepancies.

The OH molecule has energy levels with high critical densities and with strong far-IR transitions to low energy levels that are favorable to IR photoexcitation.  Radiative pumping may therefore severely alter the level populations \citep[e.g.][]{Wampfler2010}.  The IR pumping would increase the populations in excited levels, which may cause us to overestimate the total number of OH molecules for the given temperature and density.  As with H$_2$O, a rigorous assessment of the IR pumping requires a full physical model of the envelope and is beyond the scope of this work.

\subsection{O Excitation}

From the 63.18 $\mu$m line flux and assuming $T=1500$ K, the total number of neutral O atoms $\log \mathcal{N}=47.9$.  This number is robust to changes in temperature and density within the parameter space discussed here, based on RADEX models of [O I] lines.  If the [O I] emission is spread out over a circle of 100 AU in radius, the column density $\log$~N(O)$=16.5$ is much less than the column density required for the line to become optically thick ($\log$~N(O)$\sim19$ for a Gaussian profile with a FWHM of $25$ \kms).    These physical properties are adopted from the H$_2$O emission for simplicity but are likely incorrect because the location of [O I] emission is spatially different than the H$_2$O emission (left panel of Fig.~\ref{fig:maps.ps}).

\begin{figure}[!t]
\includegraphics[width=90mm]{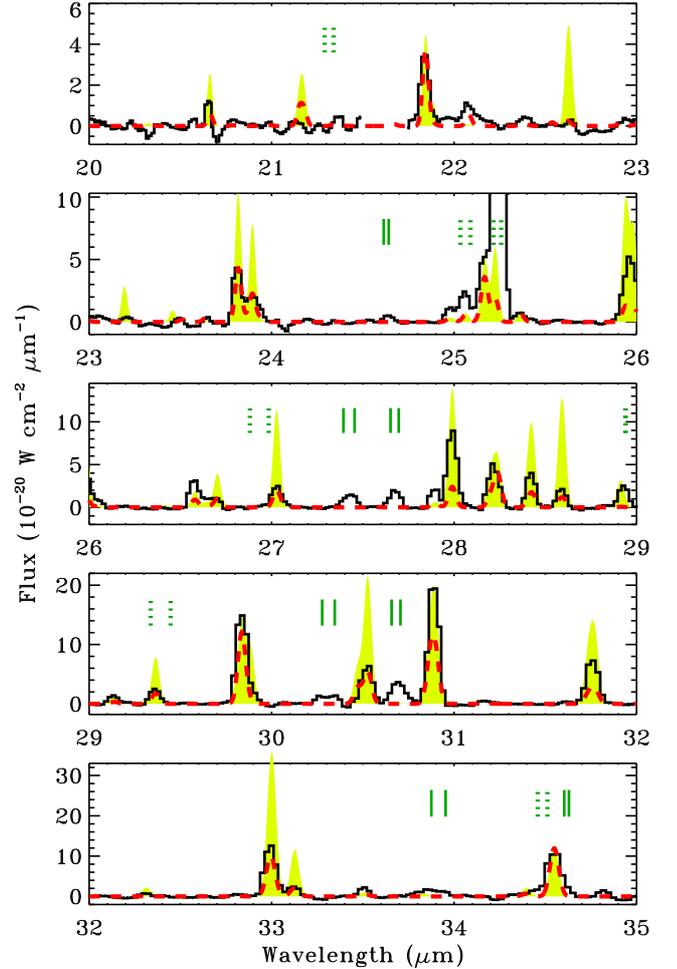}
\caption{A comparison of H12 (red dashed line, scaled to the flux in the 63.4 $\mu$m lines) and W07 (yellow filled regions, scaled to the flux of the 35.5 $\mu$m line) model spectra to the {\it Spitzer} IRS spectrum.  Green vertical lines mark the wavelengths of OH lines that are detected (solid lines) or are too weak or blended to be detected (dotted lines).  Both models reproduce most {\it Spitzer}/IRS lines reasonably well, with a few exceptions described in detail in the text.}
\label{fig:comparison.ps}
\end{figure}

\section{DISCUSSION}

\subsection{The Origin of Highly-Excited H$_2$O emission}
Prior to this work, H$_2$O emission has been attributed to three different regions in or near the IRAS 4B environment:

\vspace{1mm}

(1)  {\it A Compact Disk:}  Spectrally narrow  (FWHM$\sim 1$ \kms) H$_2^{18}$O emission is produced in a (pseudo)-disk with a radius $\sim 25$ AU around  IRAS 4B \citep{Jorgensen2010}.   From the inferred H$_2^{18}$O column density, the compact disk is optically-thick in most H$_2^{16}$O rotational transitions.  The disk covers only a small area on the sky and therefore contributes very little emission to the broad lines detected with HIFI and to the mid- and far-IR H$_2$O emission.
From the assumed $T=170$ K and resulting column density $\log N$(p-H$_2$O)$\sim18.4$ and $b=1$ \kms, the p-H$_2^{16}$O $3_{31}-2_{02}$ 138.5 $\mu$m line, from the same upper level as the observed H$_2^{18}$O line, would have a flux of $3\times10^{-22}$ W m$^{-2}$, 50 times smaller than the measured line flux.  Our PACS observations are unable to probe this disk component.

\vspace{1mm}

(2) {\it Outflow emission, (a) in low-excitation lines with a large beam:} Spectrally-broad emission in low-excitation H$_2$O lines was observed in spatially-unresolved observations over a $\sim 20-40^{\prime\prime}$ beam and was attributed to outflows based on the line widths \citep{Kristensen2010}.  The outflow is compact on the sky, so that both the red- and blue-shifted outflow lobes are located within this aperture.  { And {\it (b), in masers}: spectrally narrow ($\sim$~1~\kms) and spatially compact H$_2$O maser emission is seen from dense gas in an outflow \citep[e.g.][]{Desmurs2009}.  The relationship between the maser emission and the molecular outflow is unclear.}

\vspace{1mm}

(3) {\it Disk-envelope shock:}  H$_2$O emission in highly-excited mid-IR lines, which are spectrally and spatially-unresolved at low spectral and spatial resolution,
was attributed to the disk-envelope accretion shock \citep{Watson2007}.  The primary argument for the accretion shock is that high H$_2$ densities ($>10^{10}$ cm$^{-3}$) are needed to populate the highly excited levels, which have high critical densities.  Such high densities are expected in an envelope-disk accretion shock but are physically unrealistic for an outflow-envelope accretion shock because envelope densities are much lower than disk densities.  

\vspace{1mm}


\begin{table*}
\caption{Cooling Budget with {\it Herschel}/PACS$^a$}
{\footnotesize \begin{tabular}{|ccc|ccc|ccc|ccc|c}
\hline
Species & Obs.   &   Fraction of & \multicolumn{3}{c}{Cool Component}  &  \multicolumn{3}{c}{Warm Component} & \multicolumn{3}{c}{RADEX Model}   & \\
            & $\log$ L$^b$   & Gas Cooling$^b$   & T (K) & $\log_{10} \mathcal{N}^d$ & $\log$ L & T (K) & $\log_{10} \mathcal{N}^d$ & $\log$ L  & $T^e$ & $\mathcal{N}$ & $\log$ L  & \\
\hline
 H$_2$O      & -1.6 & 0.45  &  110  & 47.0  &-1.1  &   220 & 46.0 &-1.6  & (220)   & 48.2  &  -1.5      &  \\
 CO             & -1.6  & 0.45 & 275  & 49.6 & -1.8  &  880 & 48.5 & -1.9  & (880)   & 48.5 &  -2.0        &   \\
 OH             & -2.3  & 0.09 & 60    & 47.1 & -1.6   & 425 & 44.2 & -3.2    & (60)     & 47.6 & -2.4$^g$   & \\
{\rm [O I]$^f$}   & -3.5& 0.005   & --    & --       & --          & --   &    --  &  -- &  -- & 47.9  &  -3.5    &     \\
\hline
\multicolumn{12}{l}{$^a$All results from fits to excitation diagram except the luminosity of warm H$_2$O emission.}\\
\multicolumn{12}{l}{~~~~{ The warm and cool components are not necessarily the same for each molecule}.}\\
\multicolumn{12}{l}{~~~~{ All luminosities in units of$L_\odot$.}}\\
\multicolumn{12}{l}{$^b$PACS lines from 53--200 $\mu$m, uncorrected for extinction}\\
\multicolumn{12}{l}{$^c$RADEX model with temperature of 1500 K and $\log $n(H$_2$)$=6.5$.}\\
\multicolumn{12}{l}{$^d$Total number of emitting molecules that are detected.}\\
\multicolumn{12}{l}{$^e$The component explained by RADEX model, not the excitation temperature calculated from RADEX model.}\\
\multicolumn{12}{l}{$^f$[O I] emission is detected from a different physical location than the molecular emission.}\\
\end{tabular}}
\label{tab:cooling}
\end{table*}

Our primary goals in this work are to use the spatial distribution and excitation of warm H$_2$O emission to test (3), the envelope-disk accretion shock interpretation proposed by \citet{Watson2007}, and to subsequently use the far-IR emission to probe the heating and cooling where the emission is produced.  In the following subsections, we discuss the outflow origin of the highly-excited H$_2$O emission and subsequently discuss the implications for the envelope-disk accretion shock and outflows.

\subsection{Outflow Origin of Highly Excited H$_2$O Emission}

The H$_2$O emission detected with PACS is spatially offset from the peak of the far-IR continuum emission to the south, the direction of the blueshifted outflow.  Of the mapped lines, the H$_2$O $8_{18}-7_{07}$ and $8_{08}-7_{17}$ lines at 63.4 $\mu$m are closest in excitation to the mid-IR {\it Spitzer} lines.  The location of the emission in both 63.4 $\mu$m H$_2$O lines is consistent with the location of near-IR emission from IRAS 4B imaged with {\it Spitzer}/IRAC.  The full PACS spectrum from 50-200 $\mu$m demonstrates that all other molecular lines are also offset from the location of the sub-mm continuum peak (Fig.~\ref{fig:twospax.ps}).  These lines are produced in the southern, blueshifted outflow lobe of IRAS 4B, as shown in the plots and cartoon of Figs.~\ref{fig:maps.ps}-\ref{fig:cartoon.ps}.  { Contemporaneous to our work, \citet{Tappe2011} found that the spatial distribution of H$_2$O emission in {\it Spitzer}/IRS spectra is consistent with an outflow origin.}
The images of CO 24--23 and 49--48 also demonstrate that highly excited CO emission is produced in the blueshifted outflow lobe.  The CO 49--48 emission is located closer to the central source than the highly excited H$_2$O emission.

The combined {\it Herschel}/PACS and {\it Spitzer}/IRS excitation diagram does not show any indication of multiple components in the highly-excited H$_2$O lines.
RADEX models indicate that the highly excited PACS lines are consistent with emission from a single slab of gas with $T\sim1500$ K, $\log N$(H$_2$O)$\sim17.6$, $\log n\sim6.5$, and an emitting area equivalent to a circle with radius $\sim 100$ AU.    The same parameters reproduce the mid-IR H$_2$O emission lines detected in the {\it Spitzer}/IRS spectrum of IRAS 4B.  These same physical conditions may also produce CO fundamental emission, which could explain the bright IRAC 4.5 $\mu$m emission from the IRAS 4B outflow \citep{Tappe2011}.

The on-source component of the far-IR H$_2$O is not as well described than the outflow component because this component is faint in the short wavelength PACS lines, which  are able to constrain the properties of the emission at the offset outflow position.  The on-source emission suffers from higher extinction, so the brightest lines are at longer wavelengths, have low excitation energies, and are optically thick.   Our RADEX modeling is restricted to the higher excitation component of the H$_2$O emission.
However, the similarity of the ratio of the two 63.4 $\mu$m lines at the sub-mm continuum location and at the outflow location suggests similar excitations.  The non-detection of H$_2^{18}$O lines (see Appendix C) place a strict limit on the optical depth of the H$_2$O lines.
The H$_2$O 108.1 $\mu$m line, of which 70\% is located on-source, likely traces the same material as the HIFI spectra of low-excitation H$_2$O lines \citep{Kristensen2010}.  The width of the HIFI emission (FWHM$\sim$24 \kms, with wings that extend out to $\sim$80 \kms) is consistent with an outflow origin and inconsistent with a slow ($\sim$2 \kms) envelope disk accretion shock that would be expected for infalling gas.

In sub-mm line emission, both the red- and blueshifted outflows are detected and spatially separated (J\o rgensen et al.~2007; Yildiz et al.~submitted), which indicates that the outflow is not aligned exactly along our line of sight to the central object.  In previous near- and mid-IR imaging, only the blueshifted outflow is detected and the redshifted outflow is invisible.  The extinction to the redshifted outflow is so high that even the far-IR emission is obscured by the envelope.  The outflow therefore cannot be aligned in the plane of the sky, and likely is aligned to within $45^\circ$ of our line of sight to the central star.
  This large extinction is consistent with the outflow angle of $\sim 15^\circ$ relative to our line of sight, as inferred from assuming that the IRAS 4B outflow has the same age as the IRAS 4A outflow (Y{\i}ld{\i}z et al.~submitted).

The relationship between the outflow and the maser emission is uncertain.   
The maser emission has a position angle of 151$^\circ$ from IRAS 4B \citep{Park2007,Marvel2008,Desmurs2009}, in contrast to the $\sim174^\circ$ position angle of the molecular outflow \citep{Jorgensen2007}.  Unlike the molecular outflow, the maser emission is located in the plane of the sky ($\sim 77^\circ$ relative to our line of sight), based on the radial velocity and projected velocity on the sky.
In this case, the outflow would have a dynamical age of only 100 yr.  However, the high extinction to the redshifted outflow and
 low extinction to the blueshifted outflow together suggest that the outflow is aligned closer to our line of sight than in the plane
 of the sky.  Jet precession, perhaps a result of binarity of the central object, has been suggested as a possible explanation for the 
difference between the maser emission and molecular outflow \citep{Desmurs2009}.

H$_2$O maser emission is produced in gas with densities of $10^7-10^9$ cm$^{-3}$ \citep{Kaufman1996}.  The density of 
the far-IR H$_2$O emission could be as high as $\sim 10^{7}$ cm$^{-3}$.  However, the maser emission is produced in very 
small spots within the outflow, while the H$_2$O emission detected here is likely spread over a projected area equivalent 
to a circle with a radius $\sim 100$ AU.   In principle, the maser emission could simply be the smallest, densest 
regions within the same outflow that produces the far-IR H$_2$O emission, although the maser and far-IR H$_2$ emission may be unrelated.

{ In summary, we conclude that most of the H$_2$O emission from IRAS 4B is produced in the blueshifted outflow.  The highly excited H$_2$O emission is located at the blueshifted outflow position.  The redshifted outflow is likely not detected in these lines because any emission is obscured by the envelope.  The optically thick lower-excitation H$_2$O lines have an additional on-source component and are likely similar to the spectrally broad outflow seen by \citet{Kristensen2010}.  Similarly, the cool components of OH and CO emission are also likely produced in the outflow rather than a quiescent envelope.

\subsection{Implications for Envelope-Disk Accretion}

The primary motivation for invoking the envelope-disk accretion shock to interpret the mid-IR H$_2$O lines was the high critical densities for the mid-IR H$_2$O lines.  We have demonstrated that these lines may also be produced in gas with sub-thermal excitation.  An additional component besides the outflow does not need to be invoked at present to explain the presence of the mid-IR H$_2$O emission.

The envelope-disk accretion shock may still have some undetected contribution to the on-source H$_2$O emission 
seen with PACS, in which case the mid-IR H$_2$O emission could in principle be the combination of the blueshifted 
outflow lobe and the envelope-disk accretion shock.  The \citet{Watson2007} interpretation of the H$_2$O emission 
requires high line optical depths, including in the 63.4 $\mu$m H$_2$O lines.  However, the 63.4 $\mu$m 
H$_2$O ortho/para lines are observed at the optically-thin flux ratio, with an upper limit that no more than $30\%$ of the 
observed emission may be produced in optically thick gas.
Since the envelope-disk accretion shock would only contribute to the on-source emission, the maximum contribution 
to the total flux in the 63.4 $\mu$m lines is $\sim10\%$.  In this case, we find that the luminosity of the disk-envelope 
accretion shock would be $1.5\times10^{-3}$ L$_\odot$, 5\% of that calculated by \citet{Watson2007}$^6$.  The accretion 
rate onto the disk would correspond with $\sim 3\times10^{-6}$ $M_\odot$ yr$^{-1}$, which is likely too low to represent the main
phase of disk growth.
\footnotetext[6]{The rate is also adjusted from \citet{Watson2007} for updated distance of $235$ pc.}

An alternate and likely explanation is that any disk-envelope accretion shock is buried inside the envelope, with an extinction that prevents the detection of any mid-IR emission produced by such a shock.  The on-source H$_2$O and CO emission is likely produced in an outflow that is seen behind $A_V\sim 700$ mag. of extinction.  In this case, whether the envelope-disk accretion shock exists is uncertain and the accretion rate is completely unconstrained.  The prospects for studying envelope-disk accretion likely require a high-density ($>10^{11}$ cm$^{-3}$) tracer \citep{Blake1994} observed at high spatial and spectral resolution in the sub-mm, where emission is not affected by extinction.

\begin{table*}
\caption{Summary of components for H$_2$O emission from IRAS 4B}
\begin{tabular}{|ccccc|}
\hline
Emission source              & Probe                     & Line Profile              &  Opacity  &  Location  \\
\hline
Compact Disk         & sub-mm H$_2^{18}$O               & Narrow, symmetric & optically thick   &  on-source \\
Off-source outflow & mid/far-IR H$_2$O                  & Unresolved                              & optically thin    & off-source \\
Maser emission        & mm H$_2$O                  & Narrow                              & high density    & off-source \\
On-source outflow  &  far-IR H$_2$O             & Symmetric              & optically thin   & on-source  \\
\hline
Disk-envelope shock?   & (sub)-mm?         & Narrow (if present) & optically thick  & on-source  \\
\hline
\end{tabular}
\label{tab:components}
\end{table*}

\subsection{Gas Line Cooling Budget and Abundances for IRAS 4B}

The bulk of the far-IR line emission is produced in the outflow.  The emission is smoothly distributed within the outflow and is consistent with emission from two different positions, one located on-source and one located off-source.   

Table~\ref{tab:cooling} lists the total cooling budget attributed to the molecular and atomic emission in the far-IR, as extracted from 
the on-source position and the blueshifted outflow lobe.  Most of the cooling is in molecular lines rather than atomic lines, 
which is consistent with previous estimates of far-IR cooling from a larger sample of Class 0 objects that were observed 
with ISO \citep{Nisini2002}.  The abundance ratio$^7$ of OH/H$_2$O$\sim 0.2$
is consistent with the abundance ratio OH/H$_2$O$>0.03$ for the outflow from the high-mass 
YSO W3 IRS 5 \citep{Wampfler2011}, the only other YSO with such a measurement.  The H$_2$O/CO abundance is $\sim 1$, on the high end of the value of 0.1-1 found from emission in lower-excitation lines of CO and H$_2$O \citep{Kristensen2010}.   The ratio of H$_2$O/H$_2$ is $\sim 10^{-4}$, based on the total number of H$_2$ molecules ($\log \mathcal{N}($H$_2$)$=52.2$) calculated from the extinction corrected fluxes of pure-rotational H$_2$ lines from \citep{Tappe2011}.  The H$_2$ and H$_2$O emission trace the same projected region on the sky (right panel of Fig.~\ref{fig:maps.ps}, with the {\it Spitzer}/IRAC imaging dominated by H$_2$ emission).
\footnotetext[7]{As calculated from the highly excited H$_2$O emission and the low temperature component of OH emission, both modeled with RADEX with the same temperature and density.    It is unclear whether the warmer component of OH emission can be reproduced by these same parameters.  The OH abundance may be severely affected by infrared pumping \citep{Wampfler2010}.}

In contrast to the luminous molecular lines, the [O I] emission from IRAS 4B is surprisingly faint.  The [O I] 63~$\mu$m line is typically the brightest far-IR emission line from low-mass protostars, with an average luminosity of $10^{-3}$ L$_{bol}$ for Class 0 stars and $10^{-2}$ L$_{bol}$ for Class I stars \citep{Giannini2001,Nisini2002}.  For IRAS 4B, the [O I] flux is $10^{-3.5}$ L$_{bol}$.  The weak [O I] emission may be a signature of the youngest protostars with outflows that have not yet escaped the dense envelope, as may be the case for IRAS 4B.  Although some [O I] emission could be located behind the envelope for IRAS 4B and other Class 0 objects, the non-detection of the [O I] 145.5 $\mu$m line suggests that neglecting extinction does not lead to a serious underestimate in [O I] line lumonosity.  The PACS spectrum also does not show any evidence for high-velocity [O I] emission from the jet, unlike the more evolved embedded object HH 46 \citep{vanKempen2010}.}

The abundance ratio of O/H$_2$O is $\sim0.1$ for the detected emission in the blueshifted outflow.  The [O I] emission is predominately located alongside the outflow axis and is offset from the location of the H$_2$O emission.  Within the shock that produces the highly excited H$_2$O emission, the O/H$_2$O abundance ratio must be much lower than 0.1.  The non-detection of [C II] emission suggests that the ionization fraction in the shock is low.  At least 90\% of the gas-phase O is in H$_2$O, CO, and OH, whichis consistent with the high H$_2$O/H$_2$ ratio measured above.


Relative to other sources, the listed H$_2$O column density is $\sim 10^2-10^4$ times larger than that found from L1157 \citep{Nisini2010,Vasta2012}.  Some of this discrepancy might be attributed to methodology.  The H$_2$O column density calculated here is measured directly from the opacities in many different lines, which are spectrally unresolved.  In contrast, the H$_2$O column density is calculated from optically thick lines, many of which are spectrally resolved but measured with a range of beam sizes.  However, the H$_2$O abundance from IRAS 4B may be much larger than that of most other young stellar objects, possibly because of some physical difference in the outflow properties.

\subsection{Shock Properties for the Highly Excited Molecular Outflow}

The H$_2$O line cooling is qualitatively consistent with excitation of molecular gas predicted from C-shocks by \citet{Kaufman1996}.  The OH emission also likely traces shocks because OH has a high critical density and is an intermediate in the high temperature gas phase chemistry that produces H$_2$O.  The two-temperature shape of the CO excitation diagram is expected from models of shock and UV-excitation of outflow cavity walls \citep{Visser2011}  and is qualitatively consistent with similar observations of the Class I sources HH 46 and DK Cha (van Kempen et al.~2010ab).  The ratio of H$_2$O to CO luminosity is higher for IRAS 4B than for the Class I source HH 46 and for the borderline Class I/II source DK Cha.  These differences may indicate that shock heating of the envelope plays a more important role than UV heating during the Class 0 stage.  Indeed, \citet{Visser2011} suggested an evolutionary trend that for more massive and denser (younger) envelopes the shock heating should dominate.  The different evolutionary stages may not change the shape of the CO ladder, as the relative contributions of the mid-$J$ (15--25) and high-$J$ (25--50) CO lines are similar for DK Cha and IRAS 4B.  The highly excited H$_2$O emission may also be especially strong from IRAS 4B because the outflow covers only a very compact area when projected on the sky.

Highly excited H$_2$O and OH lines were previously detected in mid-IR {\it Spitzer}/IRS spectra of the HH 211 bow shock, which is strong enough ($\sim 200$ \kms) to dissociate molecules at the location of direct impact.  Such a high velocity shock also produces strong UV radiation (see models by Neufeld \& Dalgarno 1989 and observations by, e.g., Raymond et al.~1997 and Walter et al.~2003) that photodissociates molecules both upstream and downstream of the shock.  In Fig.~2 of Tappe et al.~(2008), many mid-IR OH emission lines are stronger than the H$_2$O lines from HH 211.  In contrast, the mid-IR H$_2$O lines are much stronger than the mid-IR OH lines from IRAS 4B.
 In addition, the OH emission from HH 211 is detected from levels with much higher excitation energies than detected here, in a
pattern that is consistent with prompt emission following production
of OH in excited levels through photodissociation of H$_2$O by far-UV radiation (presumably Lyman
$\alpha$, Tappe et al., 2008).
The atomic fine-structure lines are also brighter from the HH 211 outflow than from IRAS 4B \citep{Giannini2001,Tappe2008}.  Thus, UV radiation
likely plays a large role in both the chemistry and excitation of OH
at the HH 211 shock position, which is well separated from the YSO itself and is
outside the densest part of the envelope.

In the case of IRAS 4B, the combination of bright H$_2$O emission and faint [O I] emission suggests that the H$_2$O dissociation rate is low.  Moreover, the OH emission from IRAS 4B is not seen from the very highly excited levels detected from HH 211.  Thus, OH likely forms through the traditional route of high temperature ($>230$~K)
chemistry of reactions between O and H$_2$ to form OH.  The OH can then collide with H$_2$ to form H$_2$O.  These reactions control the oxygen chemistry in dense C-type shocks \citep[]{Draine1983,Kaufman1996}.
The balance between O, OH and H$_2$O in
well-shielded regions depends primarily on the H/H$_2$
ratio of the gas.  In dense photo-dissociation regions, these same high temperature reactions are effective in forming H$_2$O, but the strong UV field can drive H$_2$O back into O and H$_2$O \citep{Sternberg1995}.  Indeed, for the Orion Bar, the very strong UV
field leads to a higher OH/H$_2$O abundance ratio (OH/H$_2$O$>$1) than detected here
(OH/H$_2$O$\sim0.1$), in addition to emission in other diagnostics of strong UV
radiation (strong [C II] and CH$^+$ emission) that are detected from the Orion Bar \citep{Goicoechea2011} but are not detected from
IRAS 4B.  Thus, the C-shock that produces the
OH and H$_2$O emission from IRAS 4B is likely not irradiated by UV
emission.  The shocked gas may be shielded from any UV emission produced by the
central source and internal shocks within the jet.  

This shielding lends support to a C-type shock explanation rather than
molecular formation downstream of a dissociative J-type shock because
such a shock would produce UV emission (see discussion above).  A non-dissociative shock is
also consistent with the presence of H$_2$ emission from IRAS 4B \citep{Arnold2011,Tappe2011}.  
This scenario is different than that postulated for HH 46 by \citep{vanKempen2010}, where the on-source O and OH emission
was thought to be produced by a fast dissociative J-type shock based on the
different spatial extents of OH and H$_2$O.  This scenario also differs from the interpretation of \citet{Wampfler2011} that the OH and H$_2$O emission from the high-mass YSO W3 IRS 5 arises either in a J-type shock or in a 
a UV-irradiated C-shock.  While the J-type shock or UV-irradiated C-type shock is an unlikely explanation for the H$_2$O emission, the OH emission could be produced in a different location than the H$_2$O emission, as is the case for [O I].

\section{CONCLUSIONS}

We have analyzed {\it Herschel}/PACS spectral images of far-IR H$_2$O emission from the prototypical Class 0 YSO IRAS 4B.   Table 5 summarizes the different components of H$_2$O emission from IRAS 4B.  We obtained the following results:

1.)  A rich forest of highly-excited H$_2$O, OH, and CO emission lines is detected in the blueshifted outflow from IRAS 4B.  The spectrum is more line-rich than any other low-mass YSO that has been previously published.  

2.)  Nyquist-sampled spectral maps place the highly-excited 63.4 $\mu$m lines at a average distance of $5\farcs2$ (projected distance of 1130 AU) south of IRAS 4B.    The lower-excitation H$_2$O 108.1 $\mu$m line has a centroid closer to the peak of the sub-mm continuum emission and has a larger spatial extent than the 63.4 $\mu$m emission along the outflow axis.  The far-IR H$_2$O emission can be interpreted as one component located at the blueshifted outflow position and a second component at the peak position in the mass distribution (sub-mm continuum peak).  The redshifted outflow lobe is not detected in highly-excited H$_2$O emission, likely because of a high extinction ($>1500$ mag.) through to the back side of the envelope.

3.)  The highest excitation lines detected with PACS are optically-thin, indicating an ortho-to-para ratio of $\sim 3$.  RADEX models of the highly excited H$_2$O lines indicate that the emission is produced in gas described by $T\sim1500$ K, $\log n$(H$_2$)$\sim 6.5$ cm$^{-3}$, and $\log N$(H$_2$O)$\sim17.6$ cm$^{-2}$ over an emitting area equivalent to a circle with radius $\sim100$~AU.   
These same physical parameters can reproduce the mid-IR H$_2$O emission seen with {\it Spitzer}/IRS and the CO and OH emission seen with PACS.  The total mass of warm H$_2$O in the IRAS 4B outflow is about 140 times the amount of water on Earth.

4.)  From results (2) and (3), we conclude that the bulk of the far-IR H$_2$O emission is produced in outflows.  Any contribution of the envelope-disk accretion shock to highly-excited H$_2$O lines is minimal.
Moreover, at present the mid-IR H$_2$O emission does not offer any support for the presence of an envelope-disk accretion shock, as had been previously suggested by \citet{Watson2007}.   Any mid-IR H$_2$O emission produced by the envelope-disk accretion shock is likely deeply embedded in the envelope.

5.)   In the blueshifted outflow lobe over 90\% of the gas-phase O  is in H$_2$O, CO, and OH, with H$_2$O twice as abundant than CO and 10 times more abundant than OH.   The cooling budget for gas in the envelope around IRAS 4B  is dominated by H$_2$O emission. 

6.)  The H$_2$O emission traces high densities in non-dissociative C-shocks.  In contrast, much of the heating  of lower-mass envelopes occurs by energetic radiation.  The highly excited H$_2$O in the shock-heated gas is likely shielded from UV radiation produced by both the central star and the bow shock.  The OH likely forms through reactions between O and H$_2$ and provides a pathway to form H$_2$O.

{
\acknowledgements
GJH thanks Achim Tappe for interesting and extensive discussions regarding the location of H$_2$O in the {\it Spitzer}/IRS spectra and the outflow inclination, and for detailed comments on the manuscript.  GJH also thanks Tom Megeath for discussions regarding the comparison of {\it Spitzer}/IRS and {\it Herschel}/PACS spectra, and 
Mario Tafalla, Javier Goicoechea, the anonymous referee, and the editor, Malcolm Walmsley for careful reads of the manuscript and insightful comments, Per Bjerkelli, Jeong-Eun Lee, and Doug Johnstone for some useful comments, and Stefani Germanotta for help in preparing the manuscript.
We also thank Javier Goicoechea for providing us with the SPIRE spectrum of Ser SMM1 and for use of some DIGIT data to help calibrate the PACS spectrum at long wavelengths.  Astrochemistry in Leiden is supported by NOVA, by a Spinoza grant and
grant 614.001.008 from NWO, and by EU FP7 grant 238258.  The research of JKJ is supported by a Lundbeck Foundation Junior Group Leader Fellowship and by the Danish Research Council through the Centre for Star and Planet Formation.}

\clearpage
\clearpage
\pagebreak

\begin{appendix}

{ 
\section{Calibration of first and second order light longward of 190 $\mu$m}

PACS spectra are poorly calibrated in regions where light from different orders are recorded at the same physical location on the detector, especially between 97--103 $\mu$m and longward of $190$ $\mu$m.
Most photons between 97--105 $\mu$m get dispersed into the second order and contaminate the flux at $>190$ $\mu$m because of a mismatch between the grating and the filter transmission.  Similarly, the light at 97--103 $\mu$m can be contaminated by third-order emission at $\sim 69$ $\mu$m.  
As a consequence, the first order light at 97--103 $\mu$m has low S/N and that light and the light at $>190$ $\mu$m has not previously been flux calibrated.  In this appendix, we use PACS spectra of Serpens SMM 1 (Goicoechea et al., in prep.) and HD 100546 \citep{Sturm2010}, both reduced in the same method as IRAS 4B, to calibrate the first and second order emission from PACS at $\lambda>190$ $\mu$m.  The continuum emission from HD 100546 is produced by a disk and peaks (in Jy) at $\sim 60$ $\mu$m.  The continuum emission from Serpens SMM 1 is produced in an envelope and peaks at $\sim 150$ $\mu$m.

\begin{figure}[!b]
\includegraphics[width=80mm]{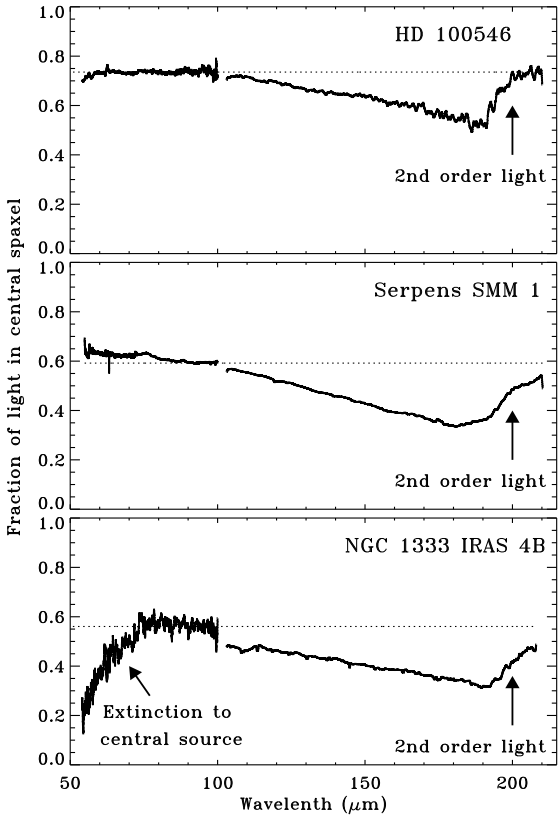}
\caption{The ratio of flux in the central spaxel to the flux in the central $3\times3$ spaxels for HD 100546, Serpens SMM 1, and NGC 1333 IRAS 4B.  The ratio falls linearly above $>100$ $\mu$m until 190 $\mu$m.  Second order light contaminates the spectrum at $>190$.  Because the first and second order light have different point-spread functions, the plotted ratio determines the fraction of first and second order photons versus wavelength.  For NGC 1333 IRAS 4B, the continuum flux in the central spaxel falls at $<70$ $\mu$m, likely because of extinction.}
\label{fig:pacspsf_iras4b.ps}
\end{figure}

First and second order light can be separated because the diffraction-limited point spread function is twice as large at 200~$\mu$m as at 100~$\mu$m.   Figure~\ref{fig:pacspsf_iras4b.ps} shows the fraction of flux in the central spaxel divided by the flux in the central  $3\times3$ spaxels.  This fraction decreases smoothly at $>100$ $\mu$m.  The fraction starts to rise at $\sim 190$ $\mu$m because of a contribution from second order emission.  At each wavelength, this fraction directly leads to a ratio of first order photons to second order photons recorded by PACS.  The second order light is then calibrated by measuring the first order light between 90--110 $\mu$m for HD 100546.  Most of the light recorded at $>190$ $\mu$m from HD 100546 is second order light. 

This calibration is then applied in reverse  to the spectrum of Serpens SMM 1 to subtract the second order emission, leaving only first order emission at $>190$ $\mu$m.  A SPIRE spectrum of Serpens SMM 1 (Goicoechea et al., in prep) is then used to flux calibrate the PACS spectrum from 190--210 $\mu$m.  Despite the red spectrum of Serpens SMM 1, $\sim 50$\% of the photons recorded at $202$ $\mu$m are second-order 101 $\mu$m photons.

To calibrate the PACS spectrum of IRAS 4B, the first and second order light are initially separated based on the point spread function for both wavelengths (Fig.~\ref{fig:pacspsf_iras4b.ps}).  The separate first and second order spectra are then flux calibrated using the relationships calculated above.  Lines are identified as first or second order emission after searching for the correct line identification at $\lambda$ and $\lambda/2$.

Figure~\ref{fig:secondord_iras4b.ps} shows the resulting spectrum of  IRAS~4B at 95--105 $\mu$m.  Analysis of a large sample of PACS spectra from the DIGIT program (P.I.~N. Evans) suggest that our flux calibration has an uncertainty of $\sim 20$\% between 98--103 $\mu$m, in addition to other sources of uncertainty in the standard PACS flux calibration.  The line fluxes measured from second order light at $>190~\mu$m are consistent to 20\% of the fluxes measured directly at 97--103 $\mu$m but with smaller error bars.  The accuracy of the flux calibration at $>190$ $\mu$m has not been evaluated but is likely uncertain to $\sim 40$\%.

The best place to observe lines between 98--103 $\mu$m with PACS is in the second order, despite overlap with first-order light because of sensitivity and higher spectral resolution.  The PACS sensitivity to first order photons at $>200$ $\mu$m is low.  This new calibration allows us to improve the accuracy of flux measurements for lines between 98--103 $\mu$m and at $>190$ $\mu$m.
}

\begin{figure}[!b]
\includegraphics[width=85mm]{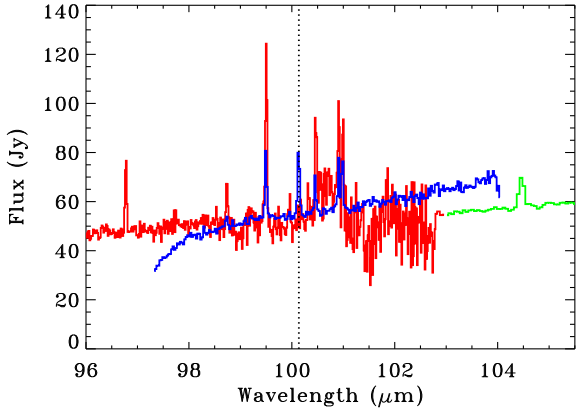}
\caption{The first and second order spectrum of IRAS 4B at $\sim 100$ $\mu$m.  The second order light at 100 $\mu$m (blue), calibrated from the $\sim 200$ $\mu$m spectrum, matches the standard first and second order spectra (red and green).  The lines in the blue spectrum are artificially weak, by a factor equivalent to the ratio of first to second order light, because the lines and continuum are treated separately.  The dotted vertical line shows the location of the CO 13--12 line at 200.27 $\mu$m.}
\label{fig:secondord_iras4b.ps}
\end{figure}

\section{Unidentified Lines}
All strong lines in the IRAS 4B spectrum are identified.  Table \ref{tab:unidentified.tab} lists several possible lines that are detected at the $\sim 3\sigma$ significance level.  The position of several of these tentative detections do not correspond to expected emission lines.  Although no o-H$_2^{18}$O are clearly detected, the o-H$_2^{18}$O $2_{12}-1_{10}$ 109.346 $\mu$m line is expected to be among the strongest H$_2^{18}$O lines and is tentatively detected.  The centroid of the detected emission is $-170$ \kms\ from the expected centroid, based on the measured wavelengths of other nearby lines.  Emission is detected consistent with the location of the
p-H$_2$O $5_{51} - 6_{24}$ 71.787 $\mu$m line, but the line flux is expected to be very weak.

These lines are all very weak and may be statistical fluctuations in the spectrum rather than significant detections of unidentified lines.

\begin{table}[!h]
\caption{Unidentified lines$^a$}
\begin{tabular}{cccccc}
\hline
$\lambda_{obs}$ ($\mu$m) & Flux$^b$ & Error$^b$ & \multicolumn{2}{c}{Possible ID} & $\lambda_{vac}$ ($\mu$m)\\
\hline
57.930   &      0.38   & 0.10  & -- & -- & --\\
60.863   &      0.27   & 0.14 & -- & -- & --\\
71.809   &       0.34    &  0.09 & p-H$_2$O & $5_{51} - 6_{24}$ & 71.787 \\
77.729   &     0.23   &  0.08    & o-H$_2$O & $7_{52}-7_{43}$ & 77.761\\
95.799   &       0.54    &  0.13  &-- & -- & --\\
109.301 &       0.39    & 0.13    & o-H$_2^{18}$O & $2_{12}-1_{10}$ & 109.346\\
112.846 &      0.24     & 0.12   & o-H$_2$O & $4_{41}-5_{14}$ & 112.802\\
127.164  &  0.21   & 0.08        & --  & -- & --\\
129.885  &      0.25    & 0.08    & -- & -- & --\\
147.380 & 0.28 & 0.08 & -- & -- & --\\
170.017 &  0.26 & 0.07 & p-H$_2$O &  $6_{33}-6_{24}$ & 170.138 \\
181.595  &      0.32   &  0.14  & -- & -- & --\\
\hline
\multicolumn{6}{l}{$^a$These lines may not be real.}\\
\multicolumn{6}{l}{$^b$$10^{-20}$ W cm$^{-2}$ s$^{-1}$.}\\
\end{tabular}
\label{tab:unidentified.tab}
\end{table}


\section{Exploring the parameter space in model fits to H$_2$O line fluxes}

\begin{figure*}[!ht]
\includegraphics[width=60mm]{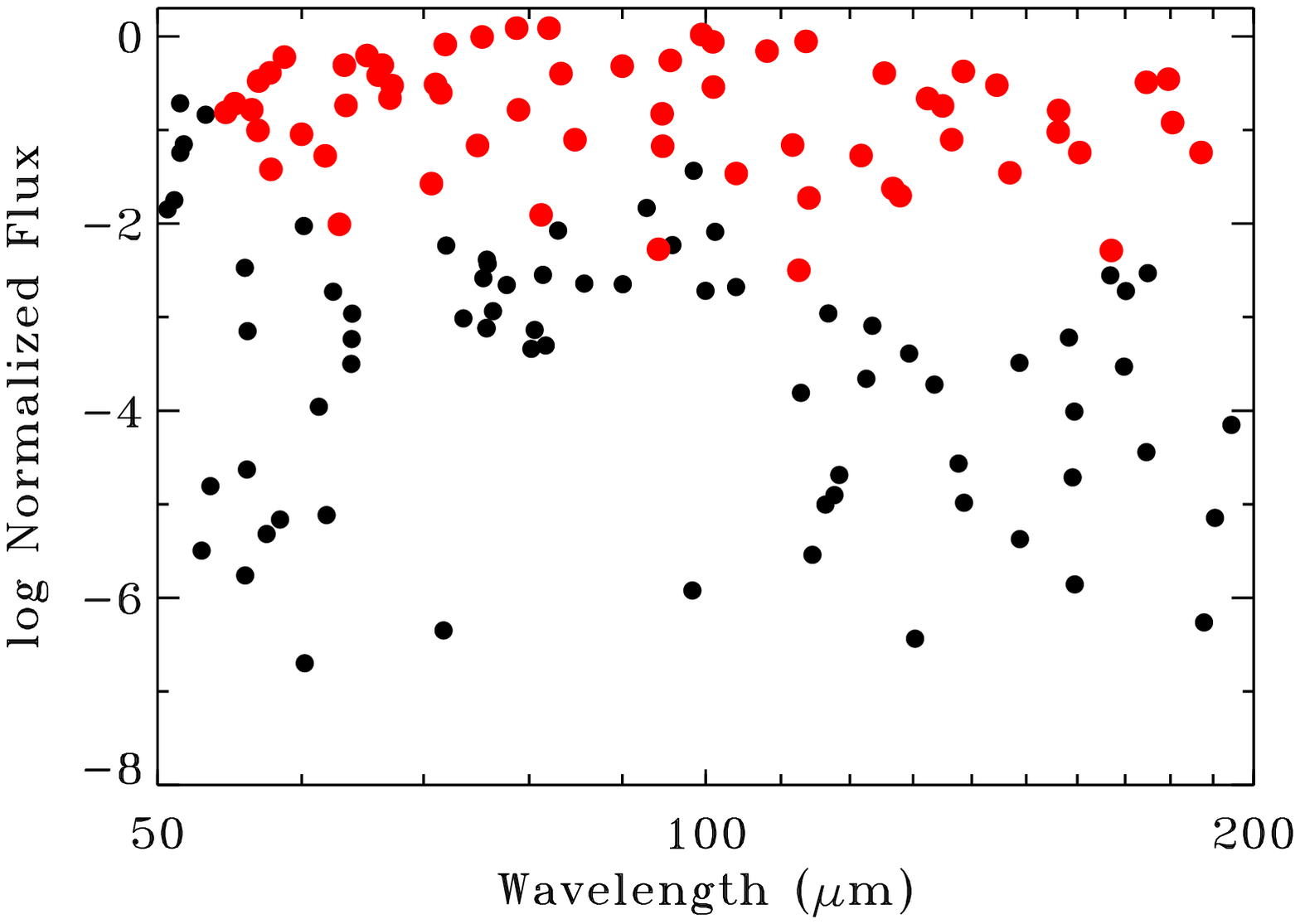}
\includegraphics[width=60mm]{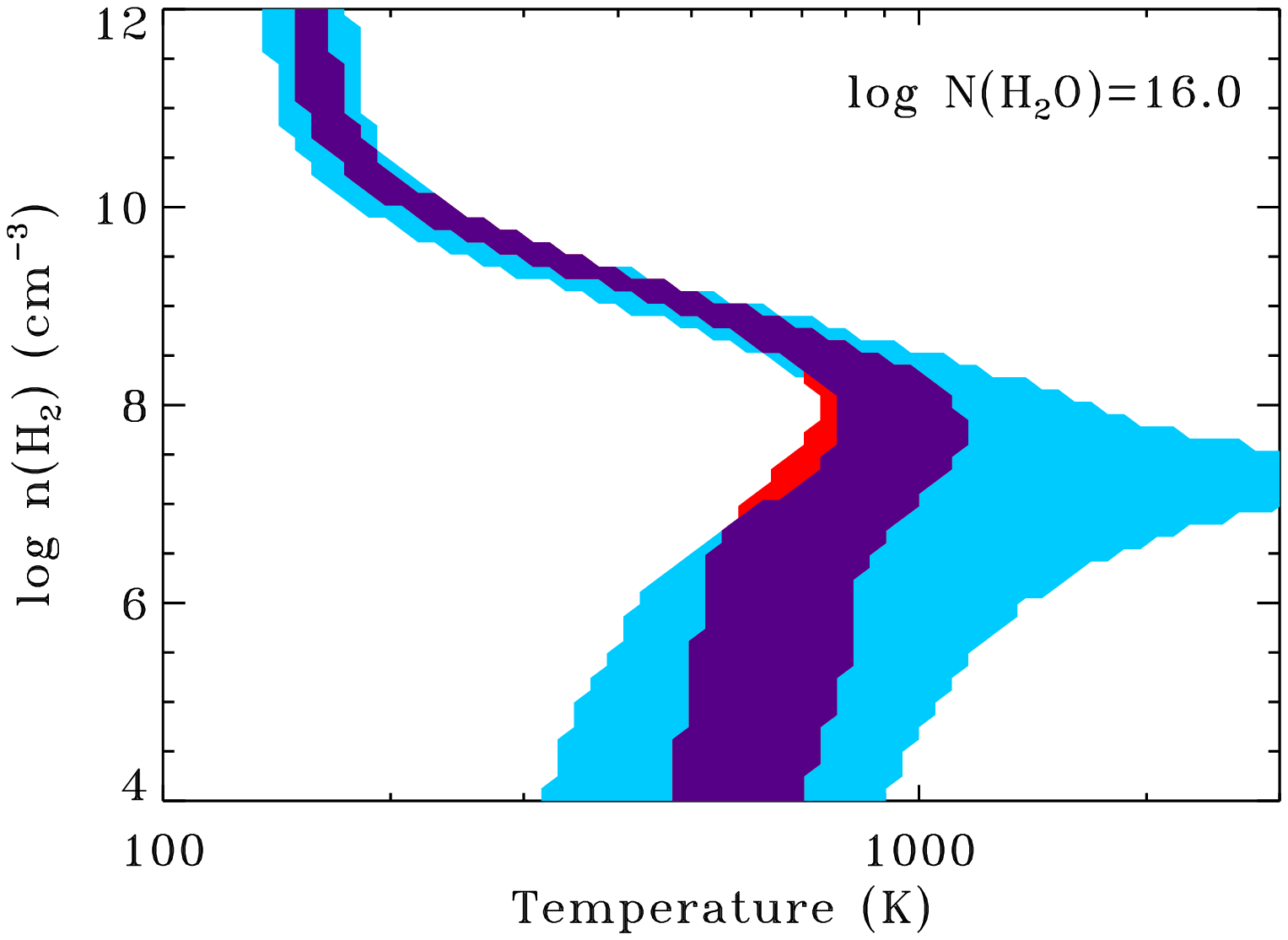}
\includegraphics[width=60mm]{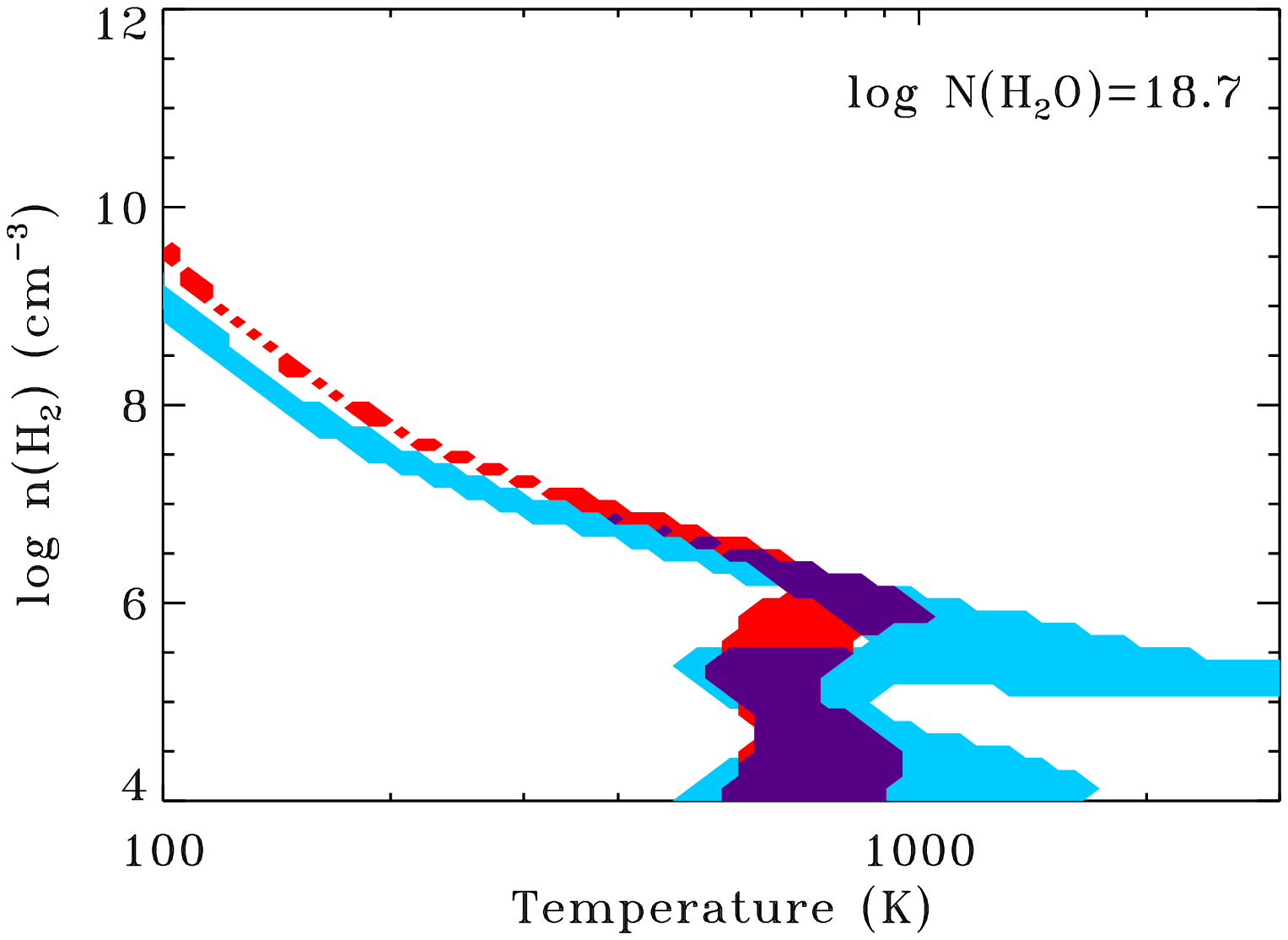}
\includegraphics[width=60mm]{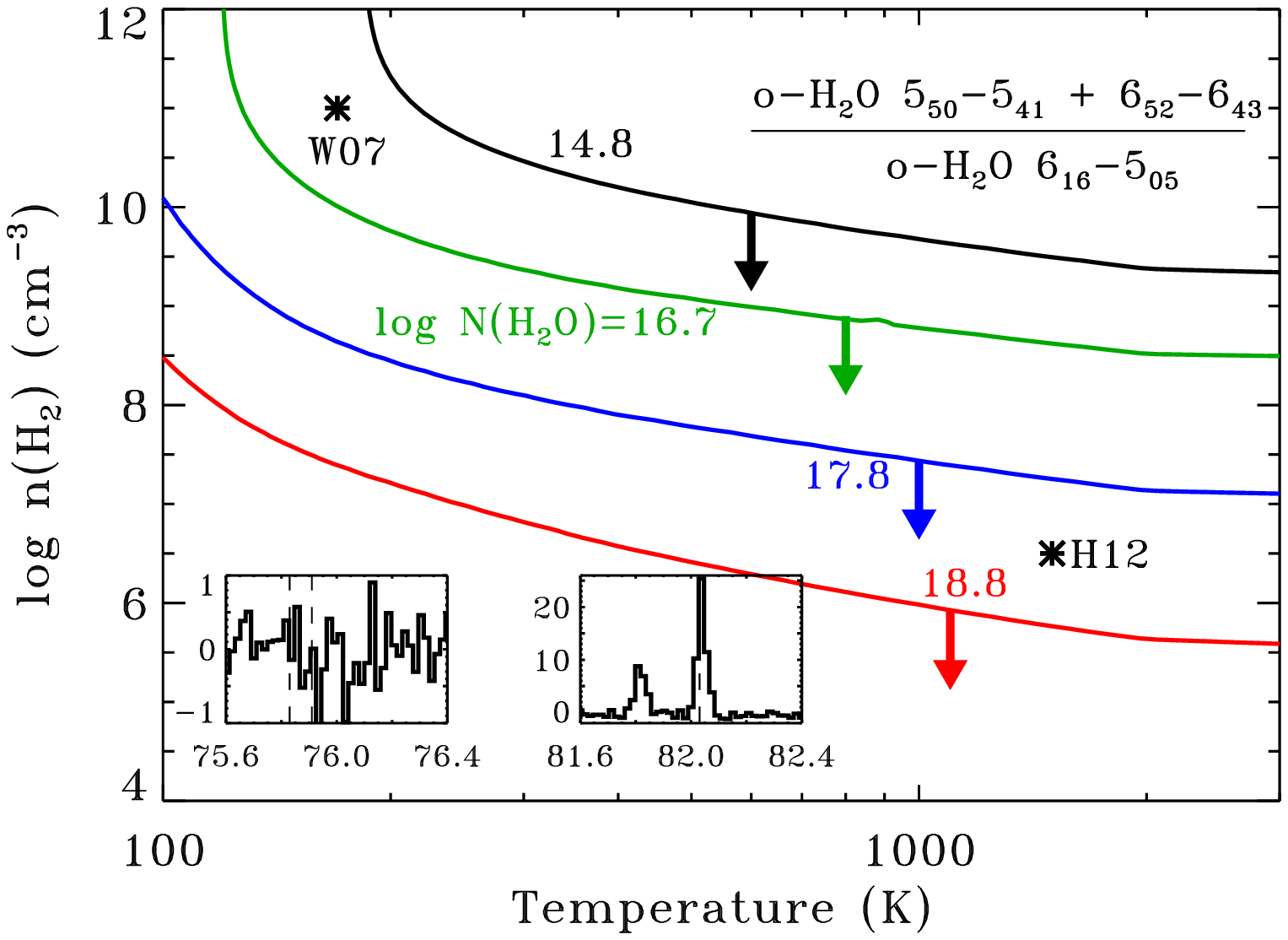}
\includegraphics[width=60mm]{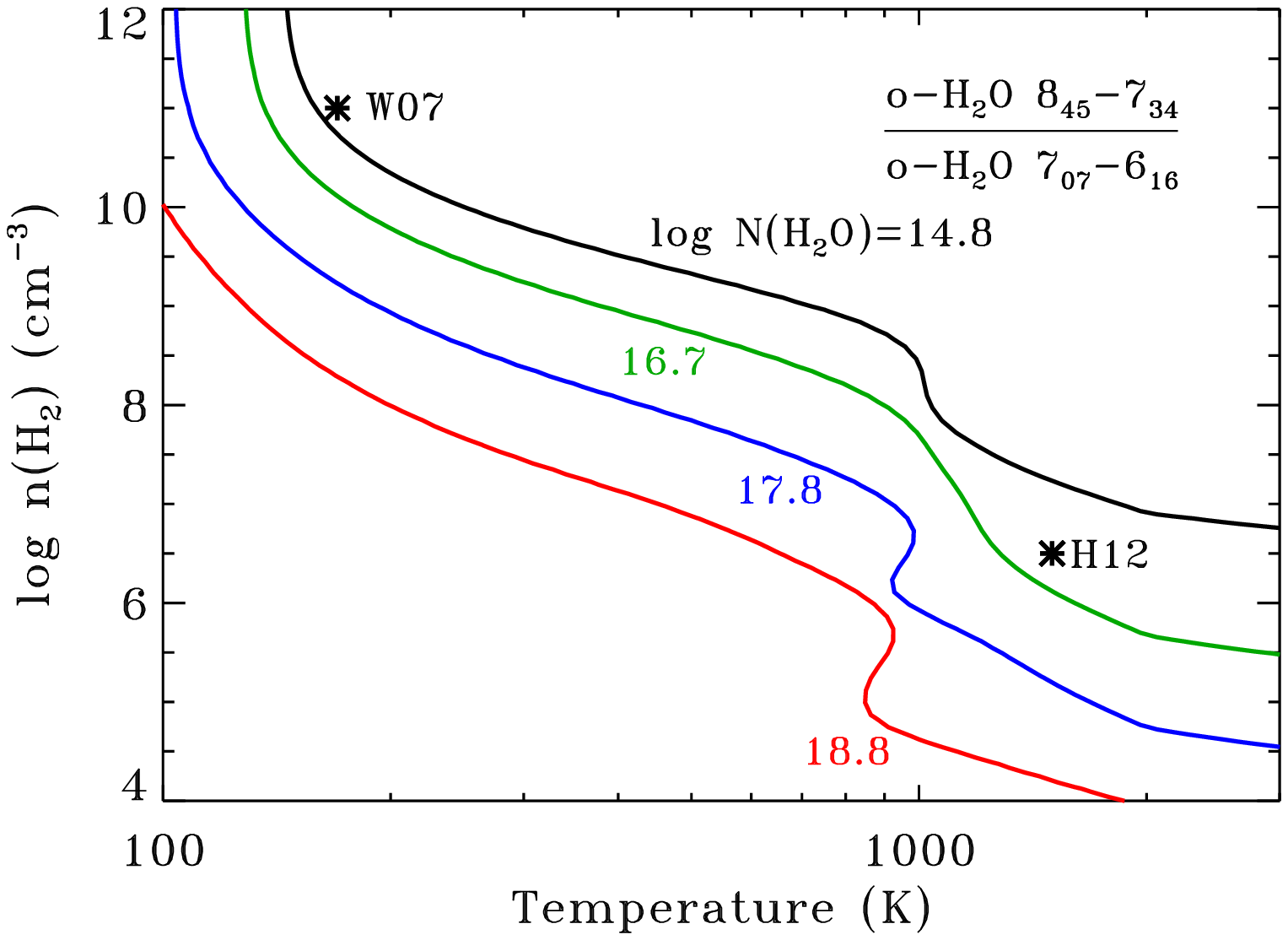}
\includegraphics[width=60mm]{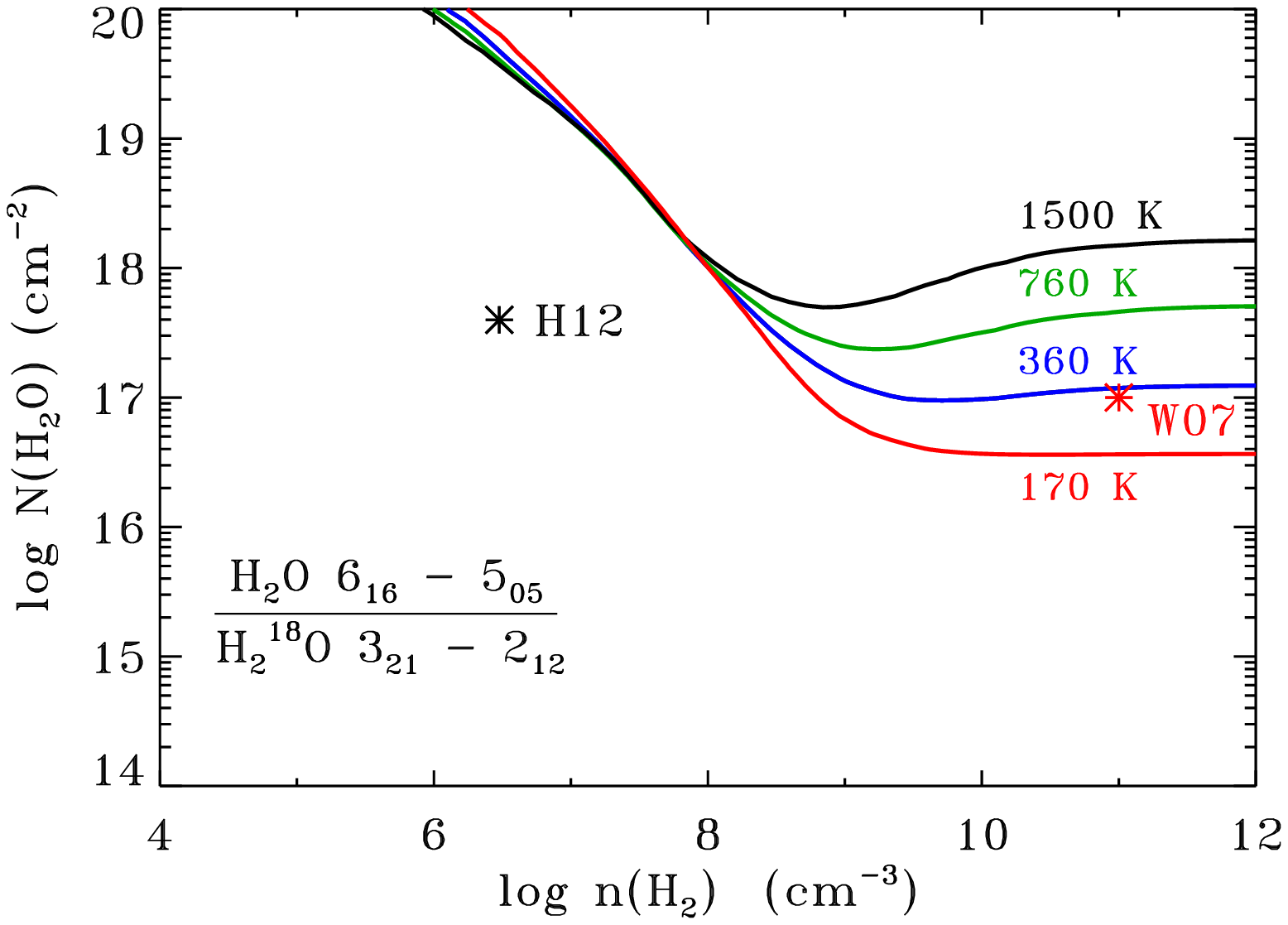}
\caption{H$_2$O fluxes and acceptable parameter space of temperature and H$_2$ density.  The upper left panel shows that in the H12 model, all far-IR H$_2$O lines expected to be strong are detected (red dots), while those expected to be weak are undetected (black dots).  Several lines at $\sim100$ $\mu$m were likely undetected because the S/N is degraded in that wavelength region.  The upper middle panel shows the acceptable contours for flux ratios of o-H$_2$O $8_{18}-7_{07}$/$7_{07}-6_{16}$ (blue),  o-H$_2$O $7_{07}-6_{16}$/$6_{16}-5_{05}$ (red), and both (purple), for two different column densities.  The acceptable parameter space shows the $1\sigma$ error bars with a 20\% relative flux calibration uncertainty.  The lower left panel shows that the non-detection of the o-H$_2$O $5_{50}-5_{41}$ and $6_{52}-6_{43}$ lines relative to o-H$_2$O $6_{16}-5_{05}$ line places a limit on the optical depth of the slab.  The lower middle panel shows where the flux ratio of o-H$_2$O $8_{45}-7_{34}$ 35.669 $\mu$m to o-H$_2$O $7_{07}-6_{16}$ 71.947 $\mu$m is equal to the observed value of 0.07.  The lower right panel shows the parameter space (high density and high column density) ruled out by the non-detection of the o-H$_2^{18}$O $3_{21}-2_{12}$ line, for four different temperatures}
\label{fig:badspace.ps}
\end{figure*}

In \S 4, we characterized the physical properties of the H$_2$O emission region by comparing the observed line fluxes to synthetic fluxes calculated in RADEX models of a plane-parallel slab.  In this appendix, we describe how the measured line fluxes constrain the parameters of the highly excited gas.

The backbone o-H$_2$O lines $9_{09}-8_{18}$, $8_{18}-7_{07}$, $7_{07}-6_{16}$, $6_{16}-5_{05}$, $5_{05}-4_{14}$ and their p-H$_2$O counterparts are amongst the best diagnostics for the excitation and optical depth of the H$_2$O emitting region.  The relative flux calibration between these ortho/para pairs should be accurate to better than 5\% because the lines are located very near each other.
 The flux ratio ranges from 2.8--3.5 in the three of the backbone line ratios with highest excitation, indicating that the  ortho-to-para ratio of $\sim 3$ is thermalized and that these lines are optically-thin.  The $6_{16}-5_{05}$, $5_{05}-4_{14}$, and $4_{14}-3_{03}$ and their p-H$_2$O counterparts have flux ratios of $\sim 2.3-2.5$, suggesting that either the modeled emission is moderately optically thick in these lines, or that a second, optically-thick component contributes some emission to the flux in lower-excitation, longer wavelength lines.  The top panels of Figure~\ref{fig:badspace.ps} show the acceptable space of n(H$_2$) and $T$ for two different values of $N$(H$_2$O) based on two of these line ratios.

The $\chi^2$ statistic automates this type of analysis over all of the applicable lines to find the best-fit parameters for a single isothermal slab.  Limiting the $\chi^2$ calculation to lines with $E^{\prime}>600$ K yields two acceptable parameter spaces, one at the H12 location ($T\sim1500$ K, $\log n$(H$_2$)$\sim6.5$, and $\log~N$(H$_2$O)$\sim17.6$) and one at low-temperature, high density, and low column density (hereafter X11, with $T\sim 200$ K, $\log n$(H$_2$)$\sim11$, and $\log~N$(H$_2$O)$\sim14.0$).  The X11 parameters differ from W07 because W07 includes a large H$_2$O column density to produce optically thick lines.  The X11 parameters are disqualified because several lines, such as o-H$_2$O $3_{30}-2_{21}$ 66.44 $\mu$m, have a synthetic flux that is much stronger than the observed emission.$^8$  As a consequence, the $\chi^2$ statistic for all lines with $E^{\prime}>400$ K yields acceptable line fluxes only around the H12 parameters, adopted in this paper, and disqualifies the X11 parameter space.
\footnotetext[8]{The opposite case, where the synthetic emission is weaker than observed in a low-excitation line, would not disqualify a model because low-excitation line emission may include an additional contribution from cooler gas.}

The o-H$_2$O transitions $5_{50}-5_{41}$ at 75.91 $\mu$m and $6_{52}-6_{43}$ at 75.83 $\mu$m have a $2\sigma$ upper limit on the combined flux of $3\times10^{-21}$ W~cm$^{-2}$.  Although these upper limits are not included in the $\chi^2$ calculations, the non-detections confirm that the parameter space with high H$_2$ density and high H$_2$O column density cannot explain the bulk of the far-IR H$_2$O emission (lower left panel of Fig.~\ref{fig:badspace.ps}).  Similarly, the line ratio o-H$_2$O $8_{45}-7_{34}$ 35.67 $\mu$m to o-H$_2$O $7_{07}-6_{16}$ 71.95 yields physical parameters consistent with the best-fit solution (lower right panel of Fig.~\ref{fig:badspace.ps}).

The non-detection of emission in H$_2^{18}$O lines (with the possible exception of o-H$_2^{18}$O $2_{12}-1_{10}$ 109.35 $\mu$m) also places a limit on the opacity of H$_2^{16}$O lines.  The bottom right panel of Fig.~\ref{fig:badspace.ps} shows where the flux ratio of o-H$_2^{18}$O $3_{21}-2_{12}$ 75.87 $\mu$m to o-H$_2$O $6_{16}-5_{05}$ 82.03 $\mu$m becomes $>50$, large enough that the H$_2^{18}$O line would be detected.  This upper limit rules out the parameter space of high density and high column density.

In the H12 model, the strength of the 6 $\mu$m continuum emission \citep{Maret2009} is similar to the strength of the 6 $\mu$m rovibrational H$_2$O lines.  The non-detection with $R=60$ spectra is marginally consistent with the predicted emission and rules out temperatures higher than $\sim 2000$ K.  An $A_V=20$ mag.~would reduce the predicted 6 $\mu$m emission by 35\%.  Decreasing the temperature from 1500 K to 1000 K would reduce the predicted 6 $\mu$m emission by a factor of $4.8$.  Measurements of emission in H$_2$O vibrational lines would place significant additional constraints on the properties of the H$_2$O emission.

\section{Effect of Extinction Estimates on Rotational Diagrams}

Extinction can affect the derivation of excitation temperatures and line luminosities, especially at shorter wavelengths where lines typically have upper levels with higher energies.
In this analysis, we assume that for all H$_2$O lines, 70\% of the luminosity is produced on source behind $A_V=700$ mag. and 30\% is produced off-source at the blueshifted outflow lobe, with $A_V=0$ mag.~(see also \S 4.3).  This description is consistent with the different observed spatial distributions of the H$_2$O 108.1 and 63.4~$\mu$m lines.   The excitation conditions and emission line ratios may instead vary smoothly with distance in the outflow.

Applying an extinction correction under these assumptions increases the excitation temperature of the warm (cool) CO component from 850 to 950 K (250 to 280 K).  If all the CO emission were located behind the $A_V$=700 mag.,  then the temperature difference would become much larger, but the short wavelength CO lines would be very faint.  The H$_2$O temperature does not change significantly because the fits are based mostly on lines at short wavelengths, so the emission from the heavily embedded region is mostly extinguished.
An $A_V=100$ mag.~to the blue outflow lobe would increase the relative luminosity of lines at $25$ $\mu$m compared with those at 100 $\mu$m by a factor of four, thereby increasing the warm H$_2$O excitation temperature from 220 K to 240 K.  An $A_V>200$ mag.~to the outflow is ruled out by the significant increase in scatter in the excitation diagram.

\clearpage

\begin{figure*}[!h]
\includegraphics[width=165mm]{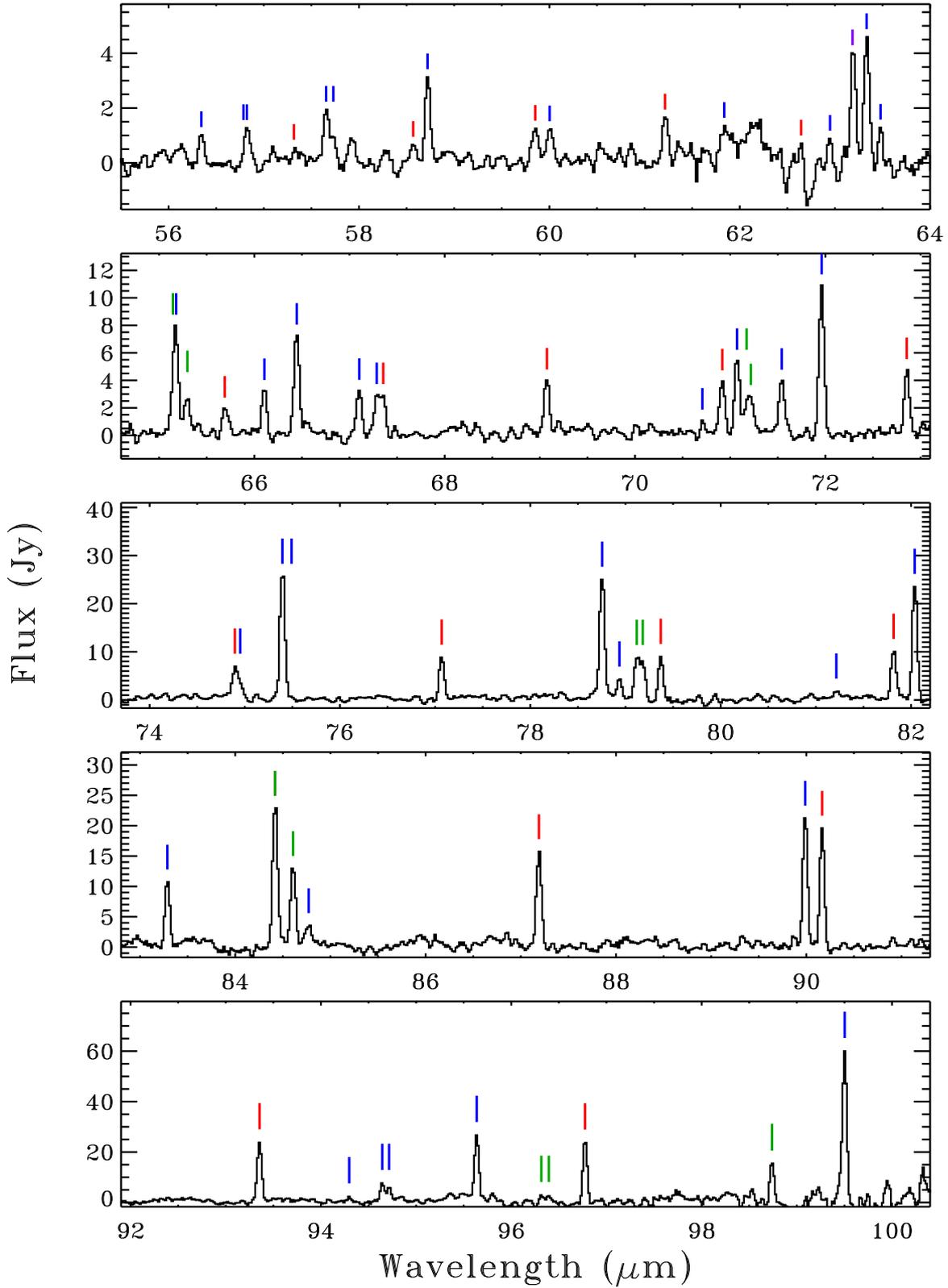}
\vspace{-15mm}
\caption{The continuum-subtracted PACS spectrum of NGC 1333 IRAS 4B from 55--100 $\mu$m.  The marks identify lines of H$_2$O (blue), CO (red), OH (green), and [O I] (purple).}
\label{fig:specall1.ps}
\end{figure*}

\begin{figure*}[!h]
\includegraphics[width=165mm]{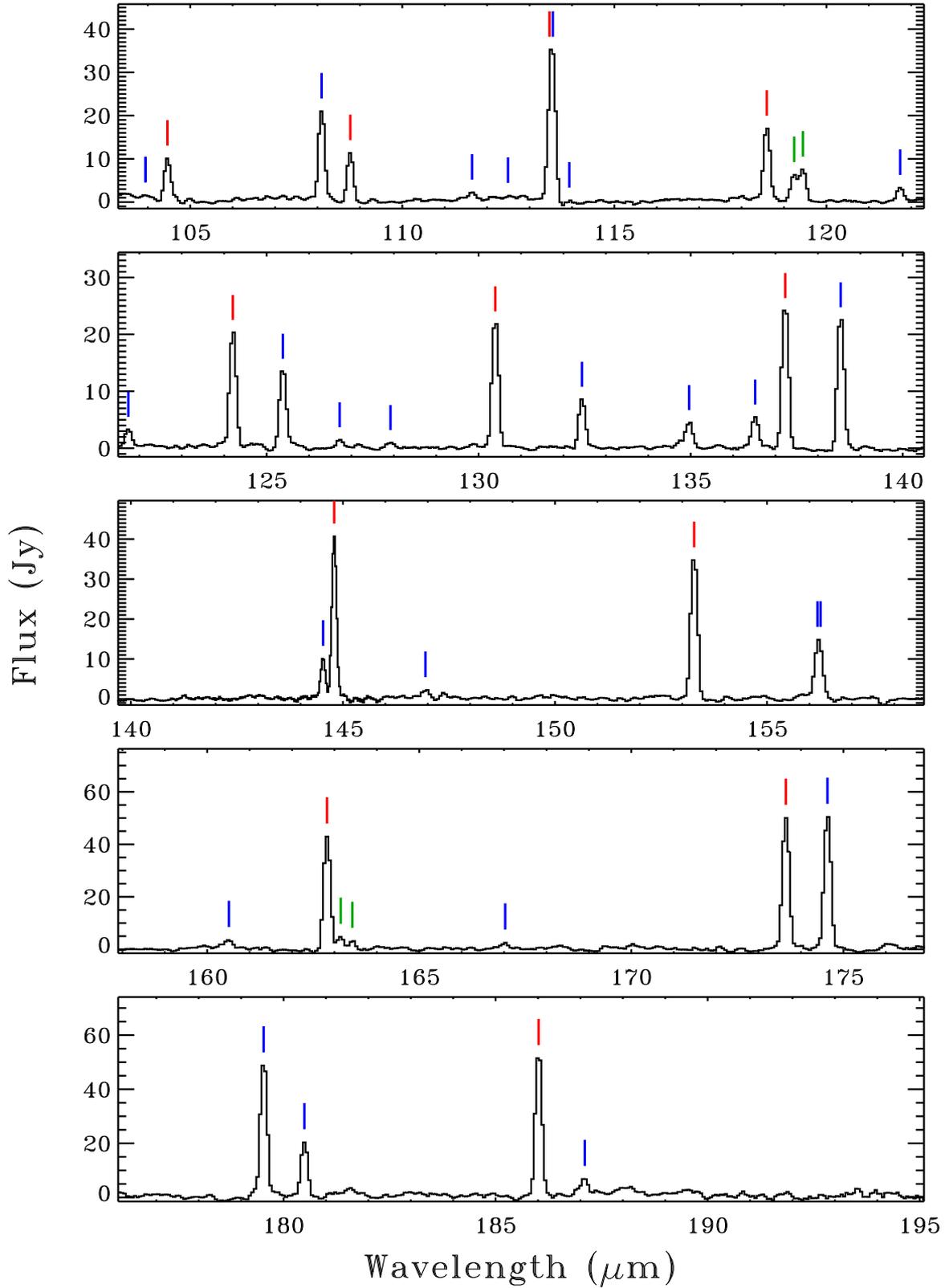}
\vspace{-15mm}
\caption{The continuum-subtracted PACS spectrum of NGC 1333 IRAS 4B from 103--195 $\mu$m.   The marks identify lines of H$_2$O (blue), CO (red), and OH (green).}
\label{fig:specall2.ps}
\end{figure*}

\clearpage
\clearpage

\begin{table*}
\caption{Detected lines in {\it Herschel}/PACS spectrum of IRAS 4B}
{\footnotesize
\begin{tabular}{cccccccc}
\hline
Species & Line & $E^\prime$ (K) & $\log A_{ul}$ (s$^{-1}$)  & $\lambda_{{\rm vac}}$ ($\mu$m) & $\lambda_{{\rm obs}}$ ($\mu$m)& Flux$^a$ & Error$^a$\\
\hline
CO & $49-48$ & 6724 & -2.17 & 53.897 & 53.893 & 0.66$^c$ & 0.11 \\
o-H$_2$O & $5_{32} - 5_{05}$ &   732   &      -1.42 &     54.506 &     54.495 &       0.61 &       0.12 \\
CO &  $48-47$ & 6457 & -2.18 & 54.986   &   54.965   &   0.39    & 0.07 \\
o-H$_2$O & $8_{27} - 7_{16}$ &  1274   &       0.28 &     55.131 &     55.146 &       0.53$^b$ &       0.13 \\
p-H$_2$O & $4_{31} - 3_{22}$ &   552   &       0.16 &     56.325 &     56.344 &       0.59 &       0.08 \\
p-H$_2$O & $9_{19} - 8_{08}$ &  1324   &       0.40 &     56.771 &     56.786 &       0.21$^b$ &       0.07 \\
o-H$_2$O & $9_{09} - 8_{18}$ &  1323   &       0.39 &     56.816 &     56.822 &       0.74 &       0.09 \\
     CO & $ 46- 45 $ &  5939 &      -2.22 &     57.308 &     57.317 &       0.27$^b$ &       0.09 \\
p-H$_2$O & $4_{22} - 3_{13}$ &   454   &      -0.42 &     57.636 &     57.655 &       0.97 &       0.11 \\
p-H$_2$O & $8_{17} - 7_{26}$ &  1270   &       0.21 &     57.709 &     57.730 &       0.43 &       0.08 \\
     CO & $ 45- 44        $ &  5688 &      -2.23 &     58.547 &     58.570 &       0.39 &       0.09 \\
o-H$_2$O & $4_{32} - 3_{21}$ &   550   &       0.14 &     58.699 &     58.721 &       1.60 &       0.09 \\
     CO & $ 44- 43        $ &  5442 &      -2.25 &     59.843 &     59.851 &       0.70 &       0.08 \\
p-H$_2$O & $7_{26} - 6_{15}$ &  1021   &       0.13 &     59.987 &     60.003 &       0.66 &       0.07 \\
     CO & $ 43- 42        $ &  4202 &      -2.27 &     61.201 &     61.215 &       0.79 &       0.10 \\
 p-H$_2$O & $4_{31}-4_{04}$ & 552 & -1.61  & 61.808 & 61.838 &  0.40  & 0.18\\
     CO & $ 42- 41        $ &  4967 &      -2.29 &     62.624 &     62.644 &       0.80$^b$ &       0.23 \\
o-H$_2$O & $9_{18} - 9_{09}$ &  1552   &      -0.38 &     62.928 &     62.949 &       0.94 &       0.18 \\
 {\rm [O I]} & $^3$P$_1-^3$P$_2$ & 228 & -4.05 & 63.185 & 63.193 & 1.83 & 0.08\\
o-H$_2$O & $8_{18} - 7_{07}$ &  1070   &       0.24 &     63.323 &     63.333 &       1.92 &       0.14 \\
p-H$_2$O & $8_{08} - 7_{17}$ &  1070   &       0.24 &     63.457 &     63.480 &       0.68 &       0.13 \\
     CO & $ 41- 40        $ &  4737 &      -2.31 &     64.117 &     64.127 &       1.08 &       0.13 \\
          OH & $^2\Pi_{3/2}~J=9/2^--7/2^+$   &  512 &   0.11 &     65.131 &     65.145 &       0.99 &       0.12 \\
o-H$_2$O & $6_{25} - 5_{14}$ &   795   &      -0.03 &     65.166 &     65.180 &       2.94 &       0.13 \\
          OH & $^2\Pi_{3/2}~J=9/2^+-7/2^-$ &  510 &       0.10 &     65.278 &     65.297 &       1.05 &       0.12 \\
     CO & $ 40- 39        $ &  4513 &      -2.34 &     65.686 &     65.689 &       0.76 &       0.10 \\
o-H$_2$O & $7_{16} - 6_{25}$ &  1013   &      -0.02 &     66.092 &     66.105 &       1.38 &       0.10 \\
o-H$_2$O & $3_{30} - 2_{21}$ &   410   &       0.09 &     66.437 &     66.446 &       3.07 &       0.13 \\
p-H$_2$O & $3_{31} - 2_{20}$ &   410   &       0.09 &     67.089 &     67.102 &       1.44 &       0.09 \\
o-H$_2$O & $3_{30} - 3_{03}$ &   410   &      -2.07 &     67.268 &     67.287 &       1.24 &       0.10 \\
     CO & $ 39- 38        $ &  4294 &      -2.36 &     67.336 &     67.355 &       1.18 &       0.10 \\
     CO & $ 38- 37        $ &  4080 &      -2.39 &     69.074 &     69.075 &       1.57 &       0.12 \\
o-H$_2$O & $8_{27} - 8_{18}$ &  1274   &      -0.48 &     70.702 &     70.707 &       0.45$^b$ &       0.12 \\
     CO & $ 37- 36        $ &  3872 &      -2.41 &     70.907 &     70.916 &       1.29 &       0.10 \\
p-H$_2$O & $5_{24} - 4_{13}$ &   598   &      -0.18 &     71.067 &     71.072 &       1.98 &       0.09 \\
          OH & $^2\Pi_{1/2}~J=7/2^--5/2^+$   &  617 &       0.01 &     71.170 &     71.172 &       0.60 &       0.07 \\
          OH &$^2\Pi_{1/2}~J=7/2^+-5/2^-$ &  617 &       0.01 &     71.215 &     71.217 &       0.75 &       0.07 \\
p-H$_2$O & $7_{17} - 6_{06}$ &   843   &       0.07 &     71.539 &     71.542 &       1.44 &       0.10 \\
o-H$_2$O & $7_{07} - 6_{16}$ &   843   &       0.06 &     71.946 &     71.961 &       3.79 &       0.12 \\
     CO & $ 36- 35        $ &  3669 &      -2.44 &     72.843 &     72.855 &       1.53 &       0.11 \\
     CO & $ 35- 34        $ &  3471 &      -2.47 &     74.890 &     74.896 &       2.01 &       0.11 \\
o-H$_2$O & $7_{25} - 6_{34}$ &  1125   &      -0.59 &     74.944 &     74.951 &       0.89 &       0.11 \\
o-H$_2$O & $3_{21} - 2_{12}$ &   305   &      -0.48 &     75.380 &     75.395 &       9.35 &       0.25 \\
o-H$_2$O & $8_{54} - 8_{45}$ &  1805   &      -0.24 &     75.495 &     75.491 &       0.33 &       0.10 \\
     CO & $ 34- 33        $ &  3279 &      -2.50 &     77.059 &     77.068 &       2.76 &       0.12 \\
o-H$_2$O & $4_{23} - 3_{12}$ &   432   &      -0.32 &     78.742 &     78.755 &       7.57 &       0.23 \\
p-H$_2$O & $6_{15} - 5_{24}$ &   781   &      -0.34 &     78.928 &     78.936 &       1.09 &       0.11 \\
          OH &$^2\Pi_{1/2}-^2\Pi_{3/2}~J=1/2^--3/2^+$ &  181 &      -1.44 &     79.115 &     79.118 &       2.53 &       0.12 \\
          OH &$^2\Pi_{1/2}-^2\Pi_{3/2}~J=1/2^+-3/2^-$ &  181 &      -1.44 &     79.178 &     79.182 &       2.16 &       0.12 \\
     CO & $ 33- 32        $ &  3092 &      -2.53 &     79.360 &     79.370 &       2.66 &       0.12 \\
p-H$_2$O & $7_{26}-7_{17}$ &  1021 & -0.61 & 81.215 & 81.215 & 0.38 & 0.07\\ 
    CO & $ 32- 31        $ &  2911 &      -2.56 &     81.806 &     81.817 &       2.64 &       0.09 \\
o-H$_2$O & $6_{16} - 5_{05}$ &   643   &      -0.13 &     82.031 &     82.038 &       6.67 &       0.16 \\
p-H$_2$O & $6_{06} - 5_{15}$ &   642   &      -0.15 &     83.283 &     83.287 &       2.75 &       0.11 \\
          OH &$^2\Pi_{3/2}~J=7/2^+-5/2^-$&  291 &      -0.28 &     84.420 &     84.418 &       5.90 &       0.22 \\
          CO &$ 31-30 $ & 2735 & -2.60 & 84.410 & \multicolumn{3}{c}{blend with OH $^2\Pi_{3/2}~J=7/2^+-5/2^-$}  \\
          OH &$^2\Pi_{3/2}~J=7/2^--5/2^+$&  290 &      -0.28 &     84.596 &     84.607 &       3.06 &       0.11 \\
o-H$_2$O & $7_{16} - 7_{07}$ &  1013   &      -0.67 &     84.766 &     84.771 &       0.76 &       0.09 \\
\hline
\multicolumn{5}{l}{$^a10^{-20}$ W cm$^{-2}$, extracted from two spaxels as discussed in \S 2.}\\
\multicolumn{5}{l}{~~~~Listed errors are $1\sigma$ and do not include calibration uncertainty.}\\
\multicolumn{5}{l}{$^b$Flux measured from outflow spaxel only.}\\
\multicolumn{5}{l}{$^c$Flux measured from Nyquist-sampled map.}\\
\end{tabular}}
\label{tab:linelist.tab}
\end{table*}

\begin{table*}
{\footnotesize
\begin{tabular}{lccccccc}
\hline
 Species & Line & $E^\prime$ (K) & $\log A_{ul}$  (s$^{-1}$)  & $\lambda_{{\rm vac}}$ ($\mu$m) & $\lambda_{{\rm obs}}$ ($\mu$m)& Flux$^a$ & Error$^a$\\
\hline
     CO & $ 30- 29        $ &  2565 &      -2.63 &     87.190 &     87.190 &       3.38 &       0.13 \\
p-H$_2$O & $3_{22} - 2_{11}$ &   296   &      -0.45 &     89.988 &     89.988 &       4.56 &       0.10 \\
     CO & $ 29- 28        $ &  2400 &      -2.67 &     90.163 &     90.165 &       4.12 &       0.10 \\
      CO & $ 28- 27        $ &  2240 &      -2.71 &     93.349 &     93.355 &       4.36 &       0.12 \\
p-H$_2$O & $5_{42}-5_{33}$ &  878  &   -0.62  & 94.209  & 94.297  &  0.32  &  0.10 \\
o-H$_2$O & $6_{25} - 6_{16}$ &   795   &      -0.76 &     94.643 &     94.645 &       1.38 &       0.11 \\
o-H$_2$O & $4_{41} - 4_{32}$ &   702   &      -0.82 &     94.704 &     94.716 &       1.00 &       0.10 \\
p-H$_2$O & $5_{15} - 4_{04}$ &   469   &      -0.35 &     95.626 &     95.637 &       4.44 &       0.12 \\
          OH &$^2\Pi_{1/2}-^2\Pi_{3/2}~J=3/2^+-5/2^-$&  270 &      -2.03 &     96.271 &     96.316 &       0.57 &       0.11 \\
          OH &$^2\Pi_{1/2}-^2\Pi_{3/2}~J=3/2^--5/2^+$&  269 &      -2.03 &     96.362 &     96.394 &       0.44 &       0.12 \\
     CO & $ 27- 26        $ &  2086 &      -2.75 &     96.773 &     96.775 &       4.28 &       0.13 \\
          OH &$^2\Pi_{1/2}~J=5/2^--3/2^+$&  415 &      -0.45 &     98.736 &     98.728$^c$ &       2.46 &       0.43 \\
          OH &$^2\Pi_{1/2}~J=5/2^+-3/2^-$&  415 &      -0.45 &     98.764 & \multicolumn{3}{c}{blend with OH $^2\Pi_{1/2}~J=5/2^--3/2^+$}\\
o-H$_2$O & $5_{05} - 4_{14}$ &   468   &      -0.41 &     99.492 &     99.496$^c$ &       9.85 &       0.26 \\
     CO & $ 26- 25        $ &  1937 &      -2.80 &    100.460 &    100.464$^c$ &       3.86  &       0.20 \\
o-H$_2$O & $5_{14} - 4_{23}$ &   574   &      -0.81 &    100.912 &    100.910$^c$ &       4.56 &       0.22 \\
p-H$_2$O & $2_{20} - 1_{11}$ &   195   &      -0.58 &    100.982 &    100.981$^c$ &       5.42 &       0.17 \\
p-H$_2$O & $6_{15}-6_{06}$ & 781  & -0.87  &  103.916 & 103.944 & 3.50 & 0.80 \\
    CO & $ 25- 24        $ &  1794 &      -2.84 &    104.445 &    104.464 &       4.22 &       0.18 \\
o-H$_2$O & $2_{21} - 1_{10}$ &   194   &      -0.59 &    108.072 &    108.097 &       9.47 &       0.19 \\
     CO & $ 24- 23        $ &  1656 &      -2.89 &    108.763 &    108.774 &       4.86 &       0.15 \\
p-H$_2$O & $5_{24} - 5_{15}$ &   598   &      -0.92 &    111.627 &    111.646 &       0.71 &       0.10 \\
o-H$_2$O & $7_{43} - 7_{34}$ &  1339  &      -0.67 &    112.510 &    112.492 &       0.21$^b$ &       0.09 \\
     CO & $ 23- 22        $ &  1524 &      -2.94 &    113.458 &    113.470 &       4.88 &       0.15 \\
o-H$_2$O & $4_{14} - 3_{03}$ &   323   &      -0.61 &    113.536 &    113.549 &      11.70 &       0.15 \\
p-H$_2$O & $5_{33} - 5_{24}$ &   725   &      -0.78 &    113.947 &    113.937 &       0.31 &       0.15 \\
     CO & $ 22- 21        $ &  1397 &      -3.00 &    118.581 &    118.595 &       6.46 &       0.11 \\
          OH &$^2\Pi_{3/2}~ J=5/2^--3/2^+$&  120 &      -0.86 &    119.233 &    119.241 &       2.23 &       0.12 \\
          OH &$^2\Pi_{3/2}~J=5/2^+-3/2^-$&  120 &      -0.86 &    119.441 &    119.447 &       2.83 &       0.11 \\
o-H$_2$O & $4_{32} - 4_{23}$ &   550   &      -0.91 &    121.721 &    121.740 &       1.17 &       0.08 \\
     CO & $ 21- 20        $ &  1276 &      -3.05 &    124.193 &    124.204 &       7.02 &       0.10 \\
p-H$_2$O & $4_{04} - 3_{13}$ &   319   &      -0.76 &    125.353 &    125.383 &       4.73 &       0.10 \\
p-H$_2$O & $3_{31} - 3_{22}$ &   410   &      -1.11 &    126.713 &    126.724 &       0.46 &       0.08 \\
o-H$_2$O & $7_{25} - 7_{16}$ &  1125   &      -0.87 &    127.883 &    127.922 &       0.30 &       0.05 \\
     CO & $ 20- 19        $ &  1160 &      -3.11 &    130.369 &    130.393 &       7.38 &       0.05 \\
o-H$_2$O & $4_{23} - 4_{14}$ &   432   &      -1.09 &    132.407 &    132.437 &       2.73 &       0.05 \\
o-H$_2$O & $5_{14} - 5_{05}$ &   574   &      -1.12 &    134.934 &    134.964 &       1.27 &       0.07 \\
o-H$_2$O & $3_{30} - 3_{21}$ &   410   &      -1.18 &    136.495 &    136.520 &       1.63 &       0.06 \\
     CO & $ 19- 18        $ &  1050 &      -3.18 &    137.196 &    137.230 &       7.66 &       0.07 \\
p-H$_2$O & $3_{13} - 2_{02}$ &   204   &      -0.90 &    138.527 &    138.536 &       6.96 &       0.07 \\
p-H$_2$O & $4_{13} - 3_{22}$ &   396   &      -1.48 &    144.517 &    144.533 &       2.04 &       0.07 \\
     CO & $ 18- 17        $ &   945 &      -3.24 &    144.784 &    144.794 &       8.13 &       0.09 \\
p-H$_2$O &   $4_{31}-4_{22}$   &  552 & -1.09 &      146.923   & 146.945 &  0.49   & 0.09\\
     CO & $ 17- 16        $ &   846 &      -3.32 &    153.267 &    153.282 &       8.87 &       0.06 \\
p-H$_2$O & $3_{22} - 3_{13}$ &   296   &      -1.28 &    156.193 &    156.190 &       2.16 &       0.06 \\
o-H$_2$O & $5_{23} - 4_{32}$ &   642   &      -1.95 &    156.264 &    156.261 &       1.55 &       0.06 \\
o-H$_2$O & $5_{32} - 5_{23}$ &   732   &      -1.09 &    160.509 &    160.513 &       0.66 &       0.07 \\
     CO & $ 16- 15        $ &   752 &      -3.39 &    162.812 &    162.825 &       9.13 &       0.20 \\
          OH &$^2\Pi_{3/2} ~J=3/2^+-1/2^-$&  270 &      -1.19 &    163.123 &    163.149 &       1.01 &       0.08 \\
          OH &$^2\Pi_{3/2} ~J=3/2^--1/2^+$&  269 &      -1.19 &    163.395 &    163.423 &       0.88 &       0.04 \\
p-H$_2$O &$6_{24}-6_{15}$ & 867 & -1.12 & 167.034 & 167.027 & 0.44  & 0.06\\ 
    CO & $ 15- 14        $ &   663 &      -3.47 &    173.631 &    173.643 &       9.45 &       0.11 \\
o-H$_2$O & $3_{03} - 2_{12}$ &   196   &      -1.30 &    174.624 &    174.625 &       9.53 &       0.11 \\
o-H$_2$O & $2_{12} - 1_{01}$ &   114   &      -1.25 &    179.525 &    179.532 &       8.86 &       0.10 \\
o-H$_2$O & $2_{21} - 2_{12}$ &   194   &      -1.51 &    180.487 &    180.491 &       3.56 &       0.09 \\
     CO & $ 14- 13        $ &   580 &      -3.56 &    185.998 &    186.014 &       8.64 &       0.11 \\
p-H$_2$O & $4_{13} - 4_{04}$ &   396   &      -1.43 &    187.109 &    187.102 &       0.93 &       0.10 \\
     CO & $ 13- 12        $ &   503 &      -3.66 &    200.271 &    200.262$^c$ &       6.33 &       0.10 \\
\hline
\multicolumn{5}{l}{$^a10^{-20}$ W cm$^{-2}$, extracted from two spaxels as discussed in \S 2.} \\
\multicolumn{5}{l}{~~~~Listed errors are $1\sigma$ and do not include calibration uncertainty..}\\
\multicolumn{5}{l}{$^b$Flux measured from outflow spaxel only.}\\
\multicolumn{5}{l}{$^c$Measured from first or second order emission at $>190$ $\mu$m.}\\
\end{tabular}}
\end{table*}

%

\end{appendix}

\end{document}